\documentclass[11pt, a4paper]{article}

\usepackage{amsmath,latexsym}
\usepackage{epsfig}
\usepackage[cp1251]{inputenc}
\usepackage[T2A]{fontenc}
\newcommand{\be}{\begin{equation}}
\newcommand{\ee}{\end{equation}}
\newcommand{\bea}{\begin{eqnarray}}
\newcommand{\eea}{\end{eqnarray}}

\newtheorem{theorem}{Theorem}

\newcommand{\benumerate}{\begin{enumerate}}
\newcommand{\eenumerate}{\end{enumerate}}

\newcommand{\bitemize}{\begin{itemize}}

\newcommand{\eitemize}{\end{itemize}}

\begin{document}

\title{Classification of  integrable two-component Hamiltonian systems of hydrodynamic type in $2+1$ dimensions}
\author{E.V. Ferapontov ${}^{1}$, A.V. Odesskii ${}^{2}$ and N.M.  Stoilov ${}^{1}$}
    \date{}
    \maketitle
    \vspace{-7mm}
\begin{center}
${}^{1}$ Department of Mathematical Sciences \\ Loughborough University \\
Loughborough, Leicestershire LE11 3TU \\ United Kingdom \\[2ex]
${}^{2}$ Department of Mathematics \\ Brock University \\
St. Catharines, Ontario L2S 3A1 \\ Canada \\[2ex]
e-mails: \\[1ex] \texttt{E.V.Ferapontov@lboro.ac.uk}
\\[1ex] \texttt{aodesski@brocku.ca}\\[1ex]
\texttt{N.M.Stoilov@lboro.ac.uk}
\end{center}

\bigskip

\begin{abstract}

Hamiltonian systems of hydrodynamic type  occur in a wide range of applications including fluid dynamics, the Whitham averaging procedure and the theory of Frobenius manifolds. In  $1+1$ dimensions, the requirement of the integrability  of such systems by the generalised hodograph transform implies that integrable Hamiltonians depend on a certain number of arbitrary functions of two variables. 

On the contrary, in $2+1$ dimensions the requirement of the integrability by the method of hydrodynamic reductions, which is a natural analogue of the generalised hodograph transform in higher dimensions, leads to  finite-dimensional moduli spaces of integrable Hamiltonians. In this paper we classify integrable two-component  Hamiltonian systems of hydrodynamic type for all existing classes of  differential-geometric Poisson brackets  in $2$D, establishing a parametrisation of  integrable Hamiltonians  via elliptic/hypergeometric functions.
Our approach is based on the Godunov-type representation of Hamiltonian systems, and utilises a novel construction of Godunov's systems in terms of generalised hypergeometric functions.

\bigskip

\noindent MSC: 35L40, 35L65, 37K10.

\bigskip

Keywords: Multi-dimensional Systems of  Hydrodynamic Type, Hamiltonian Structures, Dispersionless Lax Pairs, Hydrodynamic Reductions, Godunov Form, Generalised Hypergeometric Functions.
\end{abstract}

\newpage
\tableofcontents
\newpage

\section{Introduction}

A  $1+1$ dimensional quasilinear  system,
$$
{\bf u}_t=A({\bf u}){\bf u}_x,
$$
is said to be Hamiltonian if it can be   represented in the form
$
{\bf u}_t=P h_{\bf u}
$
where  $h({\bf u})$ is the Hamiltonian density, $h_{\bf u}$ is its gradient,   ${\bf u}=(u^1, \dots, u^n)$, and $P=P^{ij}$ is a  Hamiltonian operator of  differential-geometric type,
$$
P^{ij}=g^{ij}({\bf u}) \frac{d}{dx}+b^{ij}_k({\bf u})u^k_x.
$$
It was demonstrated in \cite{Dubrovin1} that
the coefficients   $g^{ij}$ define a flat contravariant metric, and  $b^i_{jk}$ can be expressed in terms of its Levi-Civita connection via
$b^{ij}_k=-g^{is}\Gamma^j_{sk}$. This implies the existence of flat coordinates where the operator $P$ assumes a constant-coefficient form,  $P^{ij}=\epsilon^i\delta^{ij}\frac{d}{dx}$. In the same coordinates,
the corresponding matrix $A$  is the Hessian matrix of  $h.$ It was conjectured by Novikov that a combination of the Hamiltonian property with the diagonalisability of the matrix $A$  implies the integrability.  This conjecture was proved by Tsarev in \cite{Tsarev0} who established the linearisability of such systems by the generalised hodograph transform. Hamiltonian systems  arise in a wide range of applications in fluid dynamics and the Whitham averaging theory. The underlying  differential-geometric structure make them the main building blocks of the theory of Frobenius manifolds \cite{Dubrovin}. Let us note that the requirement of the diagonalisability of the matrix $A$ imposes a complicated system of third order PDEs for the Hamiltonian density $h$ whose general solution is known to be parametrised by $n(n-1)/2$ arbitrary functions of two variables. Thus, the moduli space of integrable Hamiltonians in $1+1$ dimensions is essentially infinite-dimensional. We refer to \cite{ Dubrovin1, Tsarev, Dubrovin2}
for further discussion.

In this paper we provide a complete  classification of integrable  two-component $2+1$ dimensional Hamiltonian systems of hydrodynamic type. Our approach is based on the method of hydrodynamic reductions which, in some sense,  can be viewed as a natural  extension of the generalised hodograph transform to higher dimensions. Our results demonstrate that the moduli spaces of integrable Hamiltonians in $2+1$ dimensions are finite-dimensional:  the integrability conditions  for the corresponding Hamiltonian densities constitute over-determined involutive system of finite type. Modulo natural equivalence groups, this leads to finite lists of integrable Hamiltonians parametrised by elliptic/hypergeometric functions.

Thus, we consider $2+1$ dimensional quasilinear systems
\begin{equation}
{\bf u}_t=A({\bf u}){\bf u}_x+B({\bf u}){\bf u}_y
\label{2+1}
\end{equation}
which can be represented in the Hamiltonian form,
\begin{equation}
{\bf u}_t=P h_{\bf u}.
\label{Ham2+1}
\end{equation}
Here  $h({\bf u})$ is the Hamiltonian density, $h_{\bf u}$ is its gradient,  and $P=P^{ij}$ is a  Hamiltonian operator of  differential-geometric type,
$$
P^{ij}=g^{ij}({\bf u}) \frac{d}{dx}+b^{ij}_k({\bf u})u^k_x+\tilde g^{ij}({\bf u}) \frac{d}{dy}+
\tilde b^{ij}_k({\bf u})u^k_y.
$$
Operators of this form are generated by a pair of metrics $g^{ij}, \ \tilde g^{ij}$,
see \cite{Dubrovin1, Mokhov1, Mokhov2} for the general theory and classification results. The main difference from the one-dimensional situation
is that, although both metrics $g^{ij}$ and $ \tilde g^{ij}$ must necessarily be flat, they can no longer
be reduced to  constant coefficient form simultaneously: there exist obstruction tensors.
The two-component situation is understood completely. One has only three types of  Hamiltonian operators:  the first two of them can be reduced to  constant-coefficient forms,
\begin{equation}
P= \left(
\begin{array}{cc}
d/dx&0\\
0 & d/dy
\end{array}
\right)
\label{b1}
\end{equation}
and
\begin{equation}
P= \left(
\begin{array}{cc}
0&d/dx\\
d/dx & d/dy
\end{array}
\right),
\label{b2}
\end{equation}
while the third one is essentially non-constant,
\begin{equation}
P= \left(
\begin{array}{cc}2v&w\\
w & 0
\end{array}
\right)\frac{d}{dx}+
 \left(
\begin{array}{cc}
0&v\\
v & 2w
\end{array}
\right)\frac{d}{dy}+
 \left(
\begin{array}{cc}
v_x&v_y\\
w_x & w_y
\end{array}
\right).
\label{b3}
\end{equation}
Here  $v,  w$  are  components of the vector ${\bf u}$, and $h_{\bf u}=(h_v, \ h_w)^t$. Hamiltonian systems generated by the operators (\ref{b1}) - (\ref{b3}) take the form
\begin{equation}
v_t=(h_v)_x, ~~~ w_t=(h_w)_y,
\label{Hb1}
\end{equation}
\begin{equation}
v_t=(h_w)_x, ~~~ w_t=(h_v)_x+(h_w)_y,
\label{Hb2}
\end{equation}
and
\begin{equation}
v_t=(2vh_v+wh_w-h)_x+(vh_w)_y, ~~~ w_t=(wh_v)_x+(2wh_w+vh_v-h)_y,
\label{Hb3}
\end{equation}
respectively. Our main result is a description of Hamiltonian densities $h(v, w)$ for which the corresponding systems
(\ref{Hb1}) - (\ref{Hb3}) are integrable by the method of hydrodynamic reductions as proposed in \cite{Fer1}.  Let us point out that the integrability conditions for the general class of two-component systems (\ref{2+1}) were derived in \cite{Fer2} in the coordinates where the first matrix $A$ is diagonal. It was demonstrated in \cite{Odesskii0, Odesskii2} that, again in special coordinates, the
matrices $A$ and $B$ can be parametrised by  hypergeometric functions.  However, it is not clear how to isolate  Hamiltonian cases within these general descriptions. In this paper we adopt a straightforward approach and derive the integrability conditions by directly applying the method of hydrodynamic reductions to the systems (\ref{Hb1}) - (\ref{Hb3}). For Hamiltonian systems of type (\ref{Hb1}) this was done  in \cite{Fer1}. Thus,  in the Appendix we illustrate the method of hydrodynamic reductions by calculating the integrability conditions for equations of  type (\ref{Hb2}). Applied to  equations (\ref{Hb1}) - (\ref{Hb3}), this method  results in involutive systems of fourth order PDEs for the Hamiltonian densities $h(v, w)$ which  are quite complicated in general. It was observed that these systems simplify considerably under the Legendre transformation $h(v, w)\to H(V, W)$ defined as
$$
V=h_v, \ W=h_w, \ H=vh_v+wh_w-h, \ H_V=v, \ H_W=w.
$$
In the new variables, equations (\ref{Hb1}) - (\ref{Hb3}) take the so-called Godunov form,
\begin{equation}
(H_V)_t=V_x, ~~~ (H_W)_t=W_y,
\label{vw}
\end{equation}
\begin{equation}
(H_V)_t = W_x, ~~~
(H_W)_t = V_x +W_y,
\label{beg1}
\end{equation}
and
\begin{equation}
(H_V)_t=(VH_V+H)_x+(WH_V)_y, ~~~ (H_W)_t=(VH_W)_x+(WH_W+H)_y,
\label{EP}
\end{equation}
respectively. All our classification results will  be formulated in terms of the Legendre-transformed Hamiltonian densities  $H(V, W)$.
We recall that a system (\ref{2+1}) is said to be in Godunov form \cite{Godunov} if,  for appropriate potentials $F_{\alpha} ({\bf u}), \ \alpha =0, 1, 2,$ it possesses a conservative representation
$$
(F_{0, i})_t+(F_{1, i})_x+(F_{2, i})_y=0,
$$
 $F_{\alpha, i}=\partial F_{\alpha}/\partial {u^i}$. In particular, in the case (\ref{EP}) one has $F_0=-H, F_1=VH, F_2=WH$.  This representation will be utilised in Sect. 5 to provide a parametrisation of the generic integrable potential of type (\ref{EP}) by generalised hypergeometric functions.

Let us first review the known results for systems of type (\ref{vw}). Here the integrability conditions take the form
$$
\begin{array}{c}
H_{VW}H_{VVVV}=2H_{VVV}H_{VVW}, \\
\ \\
H_{VW}H_{VVVW}=2H_{VVV}H_{VWW}, \\
\ \\
H_{VW}H_{VVWW}=H_{VVW}H_{VWW}+H_{VVV}H_{WWW}, \\
\ \\
H_{VW}H_{VWWW}=2H_{VVW}H_{WWW}, \\
\ \\
H_{VW}H_{WWWW}=2H_{VWW}H_{WWW},
\end{array}
$$
see \cite{Fer1}. These equations are in involution and can be solved in closed form.  Up to the action of a natural equivalence group, this provides a complete list of integrable potentials.

\begin{theorem} \cite{FMS}
The generic integrable potential of type (\ref{vw}) is given by the formula
$$
H=Z(V+W)+\epsilon Z(V+\epsilon W)+\epsilon^2 Z(V+\epsilon^2
W)
$$
where $\epsilon=e^{2\pi i/3}$ and $Z''(s)=\zeta(s)$. Here $\zeta $ is the
Weierstrass zeta-function,  $\zeta'=-\wp, \ (\wp')^2=4\wp^3-g_3$. Degenerations of this solution  correspond to
$$
H=(V+W) \log(V+W)+\epsilon (V+\epsilon W) \log(V+\epsilon
W)+\epsilon^2 (V+\epsilon^2 W) \log(V+\epsilon^2 W),
$$
$$
H=\frac{1}{2}V^2\zeta (W), ~~~
H=\frac{V^2}{2W},
~~~
H=(V+W)\ln (V+W),
$$
as well as the following  polynomial potentials:
$$
H=V^2W^2,
~~~
H=VW^2+\frac{\alpha}{5}W^5,
~~~
H=VW+\frac{1}{6}W^3.
$$
\end{theorem}
We refer to Sect. 2 for further details.

In Sect. 3 we demonstrate that for systems of type (\ref{beg1}) the integrability conditions take the form
\bea
{\nonumber}&&H_{VVVV} = \frac{2 H^2_{VVV}}{H_{VV}}, \\
{\nonumber}&&H_{VVVW} = \frac{2 H_{VVW}H_{VVV} }{H_{VV}}, \\
{\label{leg4d}}&&H_{VVWW} = \frac{2 H^2_{VVW}}{H_{VV}}, \\
{\nonumber}&&H_{VWWW} = \frac{3 H_{VWW}H_{VVW} - H_{WWW}H_{VVV}}{H_{VV}}, \\
{\nonumber}&&H_{WWWW} = \frac{6H^2_{VWW} - 4H_{WWW}H_{VVW}}{H_{VV}}.
\eea
These equations are also in involution, and can be solved in closed form.  Up to the action of a natural equivalence group, we obtain a complete list of integrable potentials.

\begin{theorem} The generic  integrable potential  of type (\ref{beg1})  is given by the formula
$$
H = V\ln{\frac{V}{\sigma(W)}}
$$
where $\sigma$ is the Weierstrass sigma-function: $\sigma' / \sigma= \zeta,  \ \zeta' = -\wp, \ \wp^{\prime 2} = 4\wp^3 - g_3$. Its degenerations correspond to
$$
H = V\ln \frac{V}{W},
~~~
H = V\ln{V},
~~~
H = \frac{V^2}{2W} + \alpha W^7,
$$
as well as the following  polynomial  potentials:
$$
H = \frac{V^2}{2} + \frac{VW^2}{2} + \frac{W^4}{4},
~~~
H = \frac{V^2}{2} + \frac{W^3}{6}.
$$

\end{theorem}
Further details, as well as the dispersionless Lax pairs corresponding to integrable potentials from Theorem 2, are provided in Sect. 3.

\noindent {\bf Remark.} For $H = V\ln{V}$ the  equations (\ref{beg1}) take the form $V_t=VW_x, \ V_x=-W_y$, implying the Boyer-Finley equations for $V$:  $V_{xx}+(\ln V)_{ty}=0$. Similarly, the choice $H = \frac{V^2}{2} + \frac{W^3}{6}$ results in the equations
$V_t=W_x, \ V_x=WW_t-W_y$, implying the dKP equation for $W$: $W_{xx}=(WW_t-W_y)_t$. These  Hamiltonian representations have appeared previously in the literature, see e.g. \cite{B}.

The case  (\ref{EP})  turns out to be considerably more complicated.  The integrability conditions constitute an involutive system of fourth order PDEs for the potential $H$ which is not presented here due to its complexity (see Sect. 4). However, it
possesses a remarkable $SL(3, R)$-invariance which reflects the invariance of the Hamiltonian formalism (\ref{Hb3}) under  linear transformations of the independent variables $x, y, t$. Based on this invariance, we were able to classify integrable  potentials.

\begin{theorem}  The generic integrable potential $H(V, W)$ of type (\ref{EP})    is given by the series
$$
H=\frac{1}{Wg(V)}\left(1+\frac{1}{g_1(V)W^6}+\frac{1}{g_2(V)W^{12}}+... \right).
$$
Here $g(V)$ satisfies the fourth order ODE,
$$
g''''(2gg'^2-g^2g'')+2g^2g'''^2-20gg'g''g'''+16g'^3g'''+18gg''^3-18g'^2g''^2=0,
$$
whose general solution can be represented in parametric form as
$$
 g= w_2(t), ~~~ V=\frac{w_1(t)}{w_2(t)},
$$
where $w_1(t)$ and $w_2(t)$ are two linearly independent solutions of the hypergeometric equation
$$
t(1-t){d^2w}/{dt^2}-\frac{2}{9}w=0.
$$
 The coefficients $g_i(V)$ are certain explicit expressions in terms of $g(V)$, e.g., $g_1(V)=gg''-2g'^2$ and so on.
Degenerations of this solution correspond to
$$
H =  \frac{1}{Wg(V)}, ~~~
H = \frac{1}{WV},
~~~
H = V-W\log W,
~~~
H = V-W^2/2.
$$
\end{theorem}
We refer to Sect. 4 for further details.

In Sect. 5 we provide a parametrisation of the generic integrable potential $H(V, W)$  by generalised hypergeometric functions:

\begin{theorem}
 The generic integrable potential $H(V, W)$ of type (\ref{EP}) can be parametrised by generalised hypergeometric functions,
\begin{equation}
~~ H=G\Big(\frac{u_1-1}{u_2-1}\Big),  ~~ V=\frac{h_1}{h_0}, ~~ W=\frac{h_2}{h_0}.
\label{HypO}
\end{equation}
Here $G^{\prime}(t)=\frac{1}{t^{2/3}(t-1)^{2/3}}$ and $h_0, h_1, h_2$ are three linearly independent solution of the hypergeometric system
$$
u_1(1-u_1)h_{u_1,u_1} - \frac{4}{3}u_1h_{u_1}-\frac{2}{9}h= \frac{2}{3}\frac{u_1(u_1-1)}{u_1-u_2}h_{u_1} +\frac{2}{3}\frac{u_2 (u_2-1)}{u_2-u_1} h_{u_2},
$$
$$
u_2(1-u_2)h_{u_2,u_2} - \frac{4}{3}u_2h_{u_2}-\frac{2}{9}h= \frac{2}{3}\frac{u_1(u_1-1)}{u_1-u_2}h_{u_1} +\frac{2}{3}\frac{u_2 (u_2-1)}{u_2-u_1} h_{u_2},
$$
$$
h_{u_1,u_2} =  \frac{2}{3}\frac{h_{u_2}-h_{u_1}}{u_2-u_1},
$$
which can be viewed as a natural two-dimensional generalisation of the hypergeometric equation $t(1-t){d^2w}/{dt^2}-\frac{2}{9}w=0.$
\end{theorem}
This parametrisation is based on a novel  construction of integrable quasilinear systems in Godunov's form in terms of  generalised hypergeometric functions which builds on  \cite{Odesskii0, Odesskii2}. We believe that this construction is of independent interest, and can be applied to a whole variety of similar classification problems. The details are provided in Sect. 5 (Theorem 5).

Our results lead to the following general observations:

 \noindent (i) The moduli spaces of integrable Hamiltonians  in $2D$ are finite dimensional.

 \noindent (ii) In the two-component case, the actions of the natural equivalence groups on the moduli spaces of integrable Hamiltonians possess  open orbits. This leads to a remarkable conclusion that for any two-component Poisson bracket in 2D there exists a unique `generic' integrable Hamiltonian, all other cases can be obtained as its appropriate degenerations.


 We anticipate that the results of this paper will find applications in the theory of infinite-dimensional Frobenius manifolds, see \cite{Carlet}  for the first steps in this direction.

\section{Hamiltonian systems of type (\ref{Hb1})}

In this section we review the classification of  integrable systems of  the form (\ref{Hb1}),
$$
\left(
  \begin{array}{c}
  v \\
    w \\
  \end{array}
\right)_t    =
\left(
  \begin{array}{cc}
    d/dx & 0 \\
    0 & d/dy \\
  \end{array}
\right) \left(
                                            \begin{array}{c}
                                              h_v \\
                                              h_w \\
                                            \end{array}
                                          \right),
$$
or,  explicitly,
$$
v_t=(h_v)_x, ~~~ w_t=(h_w)_y.
$$
The integrability conditions were first derived in \cite{Fer1} based on the method of hydrodynamic reductions. These conditions constitute a system of fourth order PDEs for the Hamiltonian density $h(v, w)$,
\begin{equation}
\begin{array}{c}
h_{vw}(h_{vw}^2-h_{vv}h_{ww})h_{vvvv}=4h_{vw}h_{vvv}(h_{vw}h_{vvw}-h_{vv}h_{vww})

\\
+3h_{vv}h_{vw}h^2_{vvw}-2h_{vv}h_{ww}h_{vvv}h_{vvw}-h_{vw}h_{ww}h^2_{vvv}, \\
\ \\
h_{vw}(h_{vw}^2-h_{vv}h_{ww})h_{vvvw}=-h_{vw}h_{vvv}(h_{vv}h_{www}
+h_{ww}h_{vvw})\\
+3h^2_{vw}h^2_{vvw}-2h_{vv}h_{ww}h_{vvv}h_{vww}+h^2_{vw}h_{vvv}h_{vww}, \\
\ \\
h_{vw}(h_{vw}^2-h_{vv}h_{ww})h_{vvww}=4h^2_{vw}h_{vvw}h_{vww} \\
-h_{vv}h_{vvw}(h_{vw}h_{www}+h_{ww}h_{vww})-h_{ww}h_{vvv}(h_{vw}h_{vww}+h_{vv}h_{www}),

\\
\ \\
h_{vw}(h_{vw}^2-h_{vv}h_{ww})h_{vwww}=-h_{vw}h_{www}(h_{ww}h_{vvv}+h_{vv}h_{vww})

\\
+3h^2_{vw}h^2_{vww}-2h_{vv}h_{ww}h_{www}h_{vvw}+h^2_{vw}h_{www}h_{vvw}, \\
\ \\
h_{vw}(h_{vw}^2-h_{vv}h_{ww})h_{wwww}=4h_{vw}h_{www}(h_{vw}h_{vww}-h_{ww}h_{vvw})

\\
+3h_{ww}h_{vw}h^2_{vww}-2h_{vv}h_{ww}h_{www}h_{vww}-h_{vw}h_{vv}h^2_{www}. \\
\end{array}
\label{4int}
\end{equation}
This system is in involution, and its solution space is $10$-dimensional. Eqs. (\ref{4int})  can be represented in compact form as
$$
s d^4h=d^3h ds+3h_{vw}(2dh_vdh_w-h_{vw}d^2h) \det (dn)
$$
where $s=h_{vw}^2(h_{vv}h_{ww}-h_{vw}^2)$, $d^kh$  denotes $k$-th symmetric differential of $h$, and $n$ is the Hessian matrix of $h$.
The contact symmetry group of  the system (\ref{4int}) is also $10$-dimensional, consisting of the $8$-parameter  group of Lie-point symmetries,
\bea
&&{\nonumber}v \rightarrow av +  b,  \\
&&{\nonumber}w \rightarrow  cw + d,  \\
&&{\nonumber}h \rightarrow \alpha h + \beta v + \gamma w +\delta,
\eea
along with the two purely contact infinitesimal generators,
$$
-{\Omega}_{h_v}\frac{\partial}{\partial v}- {\Omega}_{ h_w}\frac{\partial}{\partial w}+
({\Omega}-h_v{\Omega}_{h_v}-h_w{\Omega}_{h_w})\frac{\partial}{\partial h}+({\Omega}_v+h_v{\Omega}_h)\frac{\partial}{\partial h_v}+({\Omega}_w+h_w{\Omega}_h)\frac{\partial}{\partial h_w},
$$
with the generating functions  $\Omega=h_v^2$ and
$\Omega=h_w^2$, respectively.
It was observed in \cite{Fer1}
that the integrability conditions  simplify under the  Legendre transformation,
$$
V=h_v, \ W=h_w, \ H=vh_v+wh_w-h, \ H_V=v, \ H_W=w,
$$
 which brings Eqs.  (\ref{Hb1}) into the Godunov form (\ref{vw}),
$$
(H_V)_t=V_x, ~~~ (H_W)_t=W_y.
$$
The integrability conditions (\ref{4int}) simplify to
\begin{equation}
\begin{array}{c}
H_{VW}H_{VVVV}=2H_{VVV}H_{VVW}, \\
\ \\
H_{VW}H_{VVVW}=2H_{VVV}H_{VWW}, \\
\ \\
H_{VW}H_{VVWW}=H_{VVW}H_{VWW}+H_{VVV}H_{WWW}, \\
\ \\
H_{VW}H_{VWWW}=2H_{VVW}H_{WWW}, \\
\ \\
H_{VW}H_{WWWW}=2H_{VWW}H_{WWW}.
\end{array}
\label{H1}
\end{equation}
This system  is also in involution, and  can be represented in compact form as
$$
Sd^4H=d^3HdS+6H_{VW} dVdW \det (dN)
$$
where $S=H_{VW}^2$, and $N$ is the Hessian matrix of $H$.
The system (\ref{H1}) is invariant under a $10$-parameter group of Lie-point symmetries,
\bea
&&{\nonumber}V \rightarrow aV+  b,  \\
&&{\nonumber}W \rightarrow  cW + d,  \\
&&{\nonumber}H \rightarrow \alpha H + \beta V^2 + \gamma W^2 +\mu V+\nu W+\delta.
\eea
Thus, both contact symmetry generators of the system (\ref{4int}) are mapped by the Legendre transformation to point symmetries of the system (\ref{H1}). One can show that the action of the symmetry group on the moduli space of solutions of the system (\ref{H1}) possesses an open orbit. The classification of integrable potentials $H(V, W)$ is performed modulo this equivalence.

\subsection{Classification of integrable potentials}

The paper \cite{FMS} provides a complete list of integrable potentials:

\medskip

\noindent {\bf Theorem 1}
{\it
The generic integrable potential of type (\ref{vw}) is given by the formula
$$
H=Z(V+W)+\epsilon Z(V+\epsilon W)+\epsilon^2 Z(V+\epsilon^2
W)
$$
where $\epsilon=e^{2\pi i/3}$ and $Z''(s)=\zeta(s)$. Here $\zeta $ is the
Weierstrass zeta-function,  $\zeta'=-\wp, \ (\wp')^2=4\wp^3-g_3$. Degenerations of this solution  correspond to
$$
H=(V+W) \log(V+W)+\epsilon (V+\epsilon W) \log(V+\epsilon
W)+\epsilon^2 (V+\epsilon^2 W) \log(V+\epsilon^2 W),
$$
$$
H=\frac{1}{2}V^2\zeta (W), ~~~
H=\frac{V^2}{2W},
~~~
H=(V+W)\ln (V+W),
$$
as well as the following  polynomial potentials:
$$
H=V^2W^2,
~~~
H=VW^2+\frac{\alpha}{5}W^5,
~~~
H=VW+\frac{1}{6}W^3.
$$}
The  corresponding systems  (\ref{vw}) possess dispersionless Lax pairs, see \cite{FMS} for further details.

\section{Hamiltonian systems of type (\ref{Hb2})}

In this section we consider  Hamiltonian systems of the form (\ref{Hb2}),
$$
\left(
  \begin{array}{c}
  v \\
    w \\
  \end{array}
\right)_t    =\left(
  \begin{array}{cc}
    0 & d/dx \\
    d/dx & d/dy \\
  \end{array}
\right)\left(
                                            \begin{array}{c}
                                              h_v \\
                                              h_w \\
                                            \end{array}
                                          \right),
$$
or, explicitly,
$$
v_t=(h_w)_x,  ~~~
 w_t=(h_v)_x+(h_w)_y.
$$
Let us first mention two well-known examples which fall into this class.

\noindent {\bf Example 1.} The Hamiltonian
\be
\nonumber h = \frac{w^2}{2} + e^v
\ee
 generates the  system
$$
v_t  = w_x , ~~~
 w_t = e^vv_x + w_y.
$$
Under the substitution $v =u_x$, $w=u_t$ these equations reduce to
\be
\nonumber u_{tt} - u_{ty} = e^{u_x}u_{xx},
\ee
which is an equivalent form of the Boyer-Finley equation \cite{BF}.

\noindent {\bf Example 2.} The Hamiltonian
\be
\nonumber h = \frac{v^2}{2} + \frac{2\sqrt{2}}{3}w\sqrt{w}
\ee
 generates the system
$$ v_t = \frac{1}{\sqrt{2w}}w_x, ~~~
w_t = v_x + \frac{1}{\sqrt{2w}}w_y.
$$
Introducing the variables $V = v, \ W = \sqrt{2w}, $  we obtain
$$
 V_t = W_x ,  ~~~
 WW_t  = V_x  + W_y.
$$
Setting $W  = u_t, \ V= u_x$ we arrive at the   dKP equation,
\be
\nonumber u_{ty}  - u_tu_{tt}  + u_{xx} = 0.
\ee

\medskip

 For the general system (\ref{Hb2}), the integrability conditions constitute a system of fourth order PDEs for the Hamiltonian density $h(v, w)$,
\begin{equation}
\begin{array}{c}
 h_{ww}(h_{vv}h_{ww}-h_{vw}^2)h_{vvvv} =
-6h_{vw}^2 h_{vvw}^2  +3 h_{ww} h_{vv} h_{vvw}^2
\\
+4 h_{vw}^2h_{vww} h_{vvv}  -  2h_{ww} h_{vw} h_{vvw} h_{vvv}+
h_{ww}^2h_{vvv}^2,\\
\  \\
h_{ww}(h_{vv}h_{ww}-h_{vw}^2)h_{vvvw} =
-3h_{vw}^2 h_{vww}h_{vvw}  + 3h_{ww} h_{vww} h_{vvw} h_{vv}
\\
- 3h_{ww} h_{vw} h_{vvw} ^2
+ h_{vw}^2 h_{vvv} h_{www} + h_{ww} h_{vw} h_{vww} h_{vvv} + h_{ww}^2 h_{vvw} h_{vvv},\\
\ \\
h_{ww}(h_{vv}h_{ww}-h_{vw}^2)h_{vvww} =
-2h_{vw}^2 h_{vww}^2 + 2h_{vv} h_{vww}^2 h_{ww}
\\
- 3h_{ww} h_{vw} h_{vww} h_{vvw} + h_{ww} h_{vv} h_{www} h_{wvv}
+ h_{ww} h_{vw} h_{www} h_{vvv} + h_{ww}^2 h_{vww} h_{vvv},\\
\ \\
h_{ww}(h_{vv}h_{ww}-h_{vw}^2)h_{vwww} =
-2h_{vw}^2 h_{www} h_{vww} -3h_{ww} h_{vww}^2 h_{vw}
\\
+ 3h_{ww} h_{vv} h_{www} h_{vww} + h_{ww} h_{vw} h_{www} h_{vvw}
+ h_{ww}^2  h_{www} h_{vvv}, \\
\ \\

h_{ww}(h_{vv}h_{ww}-h_{vw}^2)h_{wwww} =
-2h_{vw}^2 h_{www}^2  - 2h_{ww} h_{www} h_{vw}h_{vww}
\\
- 3h_{ww}^2 h_{vww}^2  + 3 h_{ww} h_{vv} h_{www}^2
+ 4h_{ww}^2 h_{www} h_{vvw}.
\end{array}
\label{4int1}
\end{equation}
As an illustration of the method of hydrodynamic reductions,  details of the derivation of these conditions are presented in the Appendix.
We have verified that  system (\ref{4int1}) is in involution, and its solution space is $10$-dimensional. The integrability conditions  can be represented in compact form as
$$
s d^4h=d^3h ds+3h_{ww}(2(dh_w)^2-h_{ww}d^2h) \det (dn)
$$
where $s=h_{ww}^2(h_{vv}h_{ww}-h_{vw}^2)$, $d^kh$ denotes $k$-th symmetric differential of $h$, and $n$ is the Hessian matrix of $h$.
This system is invariant under an $11$-dimensional group of contact symmetries which consists of the $9$-dimensional group of Lie-point symmetries,
\bea
&&{\nonumber}v \rightarrow av +  b,  \\
&&{\nonumber}w \rightarrow pv+ cw + d,  \\
&&{\nonumber}h \rightarrow \alpha h + \beta v + \gamma w +\delta,
\eea
along with the two purely contact infinitesimal generators,
$$
-{\Omega}_{h_v}\frac{\partial}{\partial v}- {\Omega}_{ h_w}\frac{\partial}{\partial w}+
({\Omega}-h_v{\Omega}_{h_v}-h_w{\Omega}_{h_w})\frac{\partial}{\partial h}+({\Omega}_v+h_v{\Omega}_h)\frac{\partial}{\partial h_v}+({\Omega}_w+h_w{\Omega}_h)\frac{\partial}{\partial h_w},
$$
with the generating functions  $\Omega=h_vh_w$ and
$\Omega=h^2_w$, respectively.
To solve this system  we  apply the Legendre transformation,
$$
V=h_v, \ W=h_w, \ H=vh_v+wh_w-h, \ H_V=v, \ H_W=w.
$$
The transformed equations (\ref{Hb2}) take the Godunov form (\ref{beg1}),
$$
(H_V)_t = W_x, ~~~
(H_W)_t = V_x +W_y,
$$
and the integrability conditions (\ref{4int1}) simplify to
\bea
{\nonumber}&&H_{VVVV} = \frac{2 H^2_{VVV}}{H_{VV}}, \\
{\nonumber}&&H_{VVVW} = \frac{2 H_{VVW}H_{VVV} }{H_{VV}}, \\
{\label{leg4d}}&&H_{VVWW} = \frac{2 H^2_{VVW}}{H_{VV}}, \\
{\nonumber}&&H_{VWWW} = \frac{3 H_{VWW}H_{VVW} - H_{WWW}H_{VVV}}{H_{VV}}, \\
{\nonumber}&&H_{WWWW} = \frac{6H^2_{VWW} - 4H_{WWW}H_{VVW}}{H_{VV}}.
\eea
This system is also in involution, and can be represented in  compact form as
$$
Sd^4H=d^3HdS-6H_{VV} (dW)^2 \det (dN),
$$
where $S=H_{VV}^2$, and $N$ is the Hessian matrix of $H$.
The system (\ref{leg4d}) is invariant under an 11-parameter  group of Lie-point symmetries,
\bea
&&{\nonumber}V \rightarrow aV + bW + p,  \\
{\label{group}}&&W \rightarrow  cW + d,  \\
&&{\nonumber}H \rightarrow \alpha H + \beta W^2 + \gamma VW + \epsilon V + \kappa W + \eta.
\eea
Thus, both contact symmetry generators of the system (\ref{4int1}) are mapped to point symmetries of the system (\ref{leg4d}). All our classification results will be performed modulo this equivalence.

\subsection{Classification of integrable potentials}

The main result of this section is  a complete  classification of solutions of the system (\ref{leg4d}):

\medskip

\noindent
{\bf{Theorem 2}}  {\it Modulo the equivalence group (\ref{group}), the  generic  integrable potential  $H(V,W)$ is given by the formula
\begin{equation}
H = V\ln{\frac{V}{\sigma(W)}},
\label{E1}
\end{equation}
where $\sigma$ is the Weierstrass sigma-function: $\sigma' / \sigma= \zeta,  \ \zeta' = -\wp, \ \wp^{\prime 2} = 4\wp^3 - g_3$. Its degenerations correspond to
\begin{equation}
H = V\ln \frac{V}{W},
\label{E2}
\end{equation}
\begin{equation}
H = V\ln{V},
\label{E3}
\end{equation}
 \begin{equation}H = \frac{V^2}{2W} + \alpha W^7, \label{E4} \end{equation}
as well as the following polynomial  potentials:
 \begin{equation}H = \frac{V^2}{2} + \frac{VW^2}{2} + \frac{W^4}{4}, \label {E5} \end{equation}
\begin{equation}H = \frac{V^2}{2} + \frac{W^3}{6}. \label{E6}
 \end{equation}}

\medskip

\centerline {\bf Proof:}

\medskip
\noindent  In the analysis of Eqs. (\ref{leg4d})  it is convenient to consider two cases,  $H_{VVV}\ne 0$ and
 $H_{VVV}= 0.$

\noindent  {\bf Case 1: $H_{VVV}\ne 0.$}  Then Eqs.  (\ref{leg4d})$_1$ and (\ref{leg4d})$_2$ imply  $H_{VVV} = cH_{VV}^2 $, $c=const$, so that  $H_{VV}=-\frac{1}{cV+f(W)}$. The substitution into (\ref{leg4d})$_3$ implies that $f(W)$ must be linear.  Modulo the equivalence group we can thus assume that  $H_{VV} = \frac{1}{V}$, which gives
$$
 H = V\ln(V) + g(W)V + h(W).
$$
The substitution of this ansatz into  (\ref{leg4d})$_4$ gives $h''' = 0$, so that we can set $h$ equal to zero modulo the equivalence group. Ultimately, the substitution into the last equation (\ref{leg4d})$_5$
 gives
 \be
\nonumber g'''' = 6g^{''2}.
\ee
The general solution of this equation is $g= - \ln{\sigma(W)}$, where $\sigma$ is the Weierstrass sigma-function defined as $(\ln\sigma)' = \zeta, \  \zeta' = -\wp$ with $\wp$ satisfying $\wp^{\prime 2} = 4\wp^3 - g_3$ (notice that $g_2=0$).  Degenerations of this solution give $g = -\ln W$ and $g = 0$.  This leads to the three cases
$$
H = V\ln{\frac{V}{\sigma(W)}}, ~~~
H = V\ln{\frac{V}{W}}, ~~~
H = V\ln{V}.
$$

\noindent {\bf Case 2: $H_{VVV}= 0.$} Then $H_{VV} = a(W)$,   and  (\ref{leg4d})$_3$ implies
\be
\nonumber a{''} = 2\frac{a'^2}{a},
\ee
so that $ a(W) = \frac{1}{pW + q}.$ There are again two subcases to consider  depending on whether $H_{VVW}$ is zero or non-zero. If $H_{VVW}\ne 0$ then $p\ne 0$ and, modulo the equivalence group,  one can set
$H_{VV} = \frac{1}{W}.$
Integrating twice with respect to $V$ we obtain
\be
\nonumber H = \frac{V^2}{2W} + b(W)V + c(W).
\ee
The substitution into (\ref{leg4d})$_4$ gives
$b''' = -3 b''/W$. Modulo the equivalence group, one can set
$ b(W) = 1/W$.  Thus we get
\be
\nonumber H = \frac{V^2}{2W} + \frac{V}{W} + c(W){\label{Hc}}.
\ee
By a translation of $V$ one can remove the intermediate term $\frac{V}{W}$, leaving $H = \frac{V^2}{2W} + c(W) $.  Then the  substitution into (\ref{leg4d})$_5$ gives
$$ c{''''} - \frac{4c{'''}}{w} = 0.
$$
Modulo the equivalence group, this gives
\be
\nonumber H = \frac{V^2}{2W} + \alpha W^7.
\ee

\noindent  In the second subcase,  $H_{VVV} = H_{VVW} = 0$,  one can normalize $H$ as $H=V^2/2+a(W) V+b(W)$. With this ansatz Eqs. (\ref{leg4d})$_1$ -(\ref{leg4d})$_3$  are trivial, while  the last two equations imply, modulo the equivalence group, that
\be
\nonumber H = \frac{V^2}{2} + \alpha VW^2 + \alpha^2 W^4 + \beta W^3.
\ee
The case $\alpha = 0$ leads to
\be
\nonumber H = \frac{V^2}{2} + \frac{W^3}{6}.
\ee
The case  $\alpha\neq  0$ leads, on appropriate rescalings, to
\be
\nonumber H = \frac{V^2}{2} + \frac{VW^2}{2} + \frac{W^4}{4}.
\ee
This finishes the proof of Theorem 2.

\subsection{Dispersionless Lax pairs}

We recall that a quasilinear system (\ref{2+1}) is said to possess a dispersionless Lax pair,
\begin{equation}
\label{Lax2} S_{x} = F\left({\bf u}, S_t \right), \qquad
S_y = G\left({\bf u}, S_t \right),
\end{equation}
 if it can be recovered from the consistency condition $S_{xy}=S_{yx}$ (we point out that the dependence of $F$ and $G$ on $S_t$ is generally non-linear). Dispersionless Lax pairs  first appeared in the construction of the universal Whitham hierarchy, see \cite{Kr} and references therein.
 It was observed in \cite{Zakharov} that  Lax pairs  of this kind naturally arise from the usual `solitonic'  Lax pairs  in the dispersionless limit.
  It was demonstrated in \cite{Fer2, Fer3} that, for a number of particularly interesting classes of systems, the existence of a dispersionless Lax pair is {\it equivalent} to the existence  of hydrodynamic reductions and, thus, to the integrability. Lax pairs  form a basis of the dispersionless $\bar \partial$-dressing method \cite{Bogdanov} and a novel version of the inverse scattering transform  associated with pararmeter-dependent vector fields \cite{Manakov}.
In this section we calculate dispersionless Lax pairs,
$$
S_x = F(V, W, S_t), ~~~ S_y  = G(V, W, S_t),
$$
for all integrable potentials appearing in Theorem 2. Thus, we require that the consistency condition $S_{xy} = S_{yx}$ results in the corresponding equations
 (\ref{beg1}). Without going into details of  calculations, which are quite standard, we only state final results. Let us  point out that in all examples discussed below the first equation of the Lax pair does not explicitly depend on $V$. The general class of Lax pairs of this kind was discussed in \cite{Odesskii1}.

 \medskip

\noindent {\bf Case (\ref{E6}):} $H = \frac{V^2}{2} + \frac{W^3}{6}.$
The  equations (\ref{beg1}) take the form
\bea
\nonumber V_t &=& W_x, \\
\nonumber WW_t &=& V_x+W_y.
\eea
The corresponding dispersionless Lax pair is
\bea
\nonumber S_x &=& \frac{S_t^2}{2} - W,  \\
\nonumber S_y &=&  - \frac{S_t^3}{3} + {S_t}W + V.
\eea

\medskip

\noindent {\bf Case (\ref{E5}):} $ H = \frac{V^2}{2} + \frac{VW^2}{2} + \frac{W^4}{4}.$
The  equations (\ref{beg1}) take the form
\bea
\nonumber V_t + WW_t &=&W_x  \\
\nonumber  VW_t
 + WV_t  + 3W^2W_t &=&V_x+W_y.
\eea
The corresponding dispersionless Lax pair is
\begin{eqnarray*}
S_x &=& \frac{S_t^4}{4} -S_tW,\\
S_y &=& - \frac{S_t^7}{7} + WS^4_t + S_t(V - W^2).
\end{eqnarray*}

\medskip

\noindent {\bf Case (\ref{E4}):} $ H = \frac{V^2}{2W}, \ \alpha=0.$
The  equations (\ref{beg1}) take the form
\bea
\nonumber  \frac{V_t}{W} - \frac{VW_t}{W^2}&=&W_x, \\
\nonumber -\frac{VV_t}{W^2} + \frac{V^2W_t}{W^3} &=& V_x+W_y.
\eea
The corresponding dispersionless Lax pair is
\bea
\nonumber S_x& =& -\frac{W^2}{2S_t^2}, \\
\nonumber S_y& =& \frac{VW}{S_t^2} + \frac{W^5}{5{S_t^5}}.
\eea

\medskip

\noindent {\bf Case (\ref{E4}):} $H = \frac{V^2}{2W} + \alpha {W^7}.$ Without any loss of generality we will set $\alpha=1/168$.
The  equations (\ref{beg1}) take the form
\bea
\nonumber  \frac{V_t}{W} - \frac{VW_t}{W^2}&=&W_x, \\
\nonumber -\frac{VV_t}{W^2} + \frac{V^2W_t}{W^3} + \frac{1}{4}W^5W_t &=& V_x+W_y.
\eea
The corresponding dispersionless Lax pair is
\bea
\nonumber S_x&=& -\frac{aW^2}{2}, \\
\nonumber S_y &=& aVW - \frac{aa'W^5}{10},
\eea
here $a=a(S_t)$ satisfies the ODE
$3a^{\prime 2} - 5 = 2aa''.$

\medskip

\noindent {\bf Case (\ref{E3}):} $H = V{\log}V.$
The  equations (\ref{beg1}) take the form
\bea
\nonumber  {V_t}&=& {V}W_x, \\
\nonumber 0&=&V_x+W_y.
\eea
The corresponding dispersionless Lax pair is
\bea
\nonumber S_x &=& -\log(W + S_t), \\
\nonumber S_y &=& \frac{V}{W + S_t} .
\eea

\medskip

\noindent {\bf Case (\ref{E2}):} $ H = V \log\frac{V}{W}.$
The  equations (\ref{beg1}) take the form
\bea
\nonumber  \frac{V_t}{V} - \frac{W_t}{W}&=&W_x, \\
\nonumber \frac{VW_t}{W^2} - \frac{V_t}{W} &=&V_x+W_y.
\eea
The corresponding dispersionless Lax pair is
\begin{eqnarray*}
S_x &=& -\ln(S_t +W) - \epsilon\ln(S_t +\epsilon{W}) - \epsilon\ln(S_t + \epsilon^2{W}), ~~~ \epsilon = e^{\frac{2\pi{i}}{3}}, \\
S_y &=& -3\frac{VWS_t}{W^3 + S_t^3}.
\end{eqnarray*}

\medskip

\noindent {\bf Case (\ref{E1}):} $H = V\log\frac{V}{\sigma(W)}$.
We recall that $(\log\sigma)' = \zeta$ and $\zeta' = - \wp$ where $\wp=\wp(w, 0, g_3)$ is the Weierstrass $\wp$-function, $\wp'^2=4\wp^3-g_3$.
The  equations (\ref{beg1}) take the form
\begin{eqnarray*}
 \frac{1}{V}V_t - \zeta(W)W_t&=&W_x,\\
 -\zeta(W)V_t + V\wp(W)W_t &=& V_x+W_y.
\end{eqnarray*}
The corresponding dispersionless Lax pair is of the form
\bea
\nonumber S_x &=& F(W,  S_t), \\
\nonumber S_y &=& V A(W,  S_t),
\eea
where the functions $F$ and $A$ satisfy the equations
$$
F_W = -A, ~~~
F_\xi =\frac{A_W}{A}  - \zeta(W)
$$
and
$$
A_\xi  = \wp(W) - \frac{A^2_W}{A^2}, ~~~
A_{WW} = 2\frac{A^2_W}{A} - 2A\wp(W),
$$
respectively (here $\xi=S_t$).
Setting $A = 1/u$ one can rewrite the last two equations for $A$ in the equivalent form,
\begin{equation}
\label{lame} u_{\xi}=u_W^2-\wp(W) u^2, ~~~ u_{WW}= 2\wp(W)u.
\end{equation}
Here the second equation is a particular case  of the  Lame equation. It has two linearly independent solutions \cite{Akhieser},
$$
e^{-W\zeta(\alpha)}\frac{\sigma(W+\alpha)}{\sigma(W) \sigma(\alpha)}, ~~~
e^{W\zeta(\alpha)}\frac{\sigma(W-\alpha)}{\sigma(W) \sigma(\alpha)},
$$
where $\alpha$ is the zero of the $\wp$-function: $\wp(\alpha)=0$.
Thus, one can set
$$
u=-e^{-W\zeta(\alpha)}\frac{\sigma(W+\alpha)}{\sigma(W) \sigma(\alpha)}\ a(\xi)+e^{W\zeta(\alpha)}\frac{\sigma(W-\alpha)}{\sigma(W) \sigma(\alpha)}\ b(\xi).
$$
The substitution into $(\ref{lame})_1$ implies a pair of ODEs for $a(\xi)$ and $b(\xi)$,
$$
a'=\wp'(\alpha) b^2, ~~~ b'=-\wp'(\alpha)a^2.
$$
This system can be solved in the form
$$
a(\xi)=c\frac{\tau \wp'(\xi, 0, g_3)-1}{\tau \wp'(\xi, 0, g_3)+1}, ~~~
b(\xi)=c\frac{\gamma \wp(\xi, 0, g_3)}{\tau \wp'(\xi, 0, g_3)+1},
$$
where the constants $c, \tau$ and $\gamma$ are defined as
$$
\tau =\frac{1}{\sqrt 3}g_3^{-1/2}, ~~~ \gamma =2g_3^{-1/3}, ~~~ c=\frac{\sqrt 3}{\wp'(\alpha)}g_3^{1/6}.
$$
Setting $g_3=1, \ \wp'(\alpha)=i$ one obtains
$$
a(\xi)=-i{\sqrt 3}\ \frac{\wp'(\xi, 0, 1)-\sqrt 3}{ \wp'(\xi, 0, 1)+\sqrt 3}, ~~~
b(\xi)=-i\ \frac{6 \wp(\xi, 0, 1)}{ \wp'(\xi, 0, 1)+\sqrt 3}.
$$
Ultimately,
$$
u=i\sqrt 3\ e^{-W\zeta(\alpha)}\frac{\sigma(W+\alpha)}{\sigma(W) \sigma(\alpha)}\  \frac{\wp'(\xi)-\sqrt 3}{ \wp'(\xi)+\sqrt 3}-6i \ e^{W\zeta(\alpha)}\frac{\sigma(W-\alpha)}{\sigma(W) \sigma(\alpha)} \ \frac{\wp(\xi)}{ \wp'(\xi)+\sqrt 3}.
$$
The functions $A$ and $F$ can be reconstructed via $A=1/u$,
$$
F_W = -1/u, ~~~
F_\xi =-\frac{u_W}{u}  - \zeta(W).
$$

\section{Hamiltonian systems of type (\ref{Hb3})}

In this section we consider  Hamiltonian systems of the form (\ref{Hb3}),
$$
\left(
  \begin{array}{c}
  v \\
    w \\
  \end{array}
\right)_t    =\Bigg [ \left(
\begin{array}{cc}2v&w\\
w & 0
\end{array}
\right)\frac{d}{dx}+
 \left(
\begin{array}{cc}
0&v\\
v & 2w
\end{array}
\right)\frac{d}{dy}+
 \left(
\begin{array}{cc}
v_x&v_y\\
w_x & w_y
\end{array}
\right)\Bigg ]\left(
                                            \begin{array}{c}
                                              h_v \\
                                              h_w \\
                                            \end{array}
                                          \right),
$$
or, explicitly,
$$
v_t=(2vh_v+wh_w-h)_x+(vh_w)_y, ~~~ w_t=(wh_v)_x+(2wh_w+vh_v-h)_y.
$$
The integrability conditions constitute a   system of fourth order PDEs for the Hamiltonian density $h(v, w)$  which is not presented here due to its complexity. We have verified that this system is in involution, and its solution space is $10$-dimensional. It is invariant under an $8$-dimensional group  of Lie-point symmetries,
\bea
&&{\nonumber}v \rightarrow av +  bw,  \\
&&{\nonumber}w \rightarrow cv+ dw ,  \\
&&{\nonumber}h \rightarrow \alpha h + \beta v + \gamma w +\delta,
\eea
along with the two purely contact infinitesimal generators,
$$
-{\Omega}_{h_v}\frac{\partial}{\partial v}- {\Omega}_{ h_w}\frac{\partial}{\partial w}+
({\Omega}-h_v{\Omega}_{h_v}-h_w{\Omega}_{h_w})\frac{\partial}{\partial h}+({\Omega}_v+h_v{\Omega}_h)\frac{\partial}{\partial h_v}+({\Omega}_w+h_w{\Omega}_h)\frac{\partial}{\partial h_w},
$$
with the generating functions  $\Omega=h_v(vh_v+wh_w)$ and
$\Omega=h_w(vh_v+wh_w)$, respectively.
As in the previous two cases, the integrability conditions for the Hamiltonian density $h(v, w)$  simplify under the Legendre transformation,
$$
V=h_v, \ W=h_w, \ H=vh_v+wh_w-h, \ H_V=v, \ H_W=w,
$$
which brings the corresponding equations (\ref{Hb3})  into the Godunov form (\ref{EP}),
$$
(H_V)_t=(VH_V+H)_x+(WH_V)_y, ~~~ (H_W)_t=(VH_W)_x+(WH_W+H)_y.
$$
\noindent {\bf Remark.} Equations (\ref{EP}) coincide with the so-called EPDiff equations \cite{Holm},
$$
{\bf m}_t-{\bf u}\times {\rm curl} \ {\bf m}+\nabla ({\bf u, m}) +{\bf m}\ {\rm div} {\bf u}=0,
$$
where ${\bf u}=-(V, W, 0)$ and ${\bf m}=(H_V, H_W, 0)$. These equations have a geometric interpretation as the Euler-Poincar\'e equations for the geodesic motion on the group of diffeomorphisms.

 The  integrability conditions   for the system (\ref{EP}) take the form
\begin{gather*}
\begin{aligned}
S H_{VVVV} =&   2 (H_{W} H_{VVV} -  H_V  H_{VVW})^2 + 4 H_V^2 (H_{VVW}^2- H_{VVV} H_{VWW}) \\
&  +12 H_W H_{VV}^2H_{VVW}  - 12 H_W  H_{VV} H_{VW}H_{VVV} -
     24 H_V  H_{VV} H_{VW} H_{VVW}\\
     &+ 8 H_V  H_{VW}^2 H_{VVV}+  6 H_V H_{VV}^2 H_{VWW} + 10 H_V  H_{VV} H_{WW} H_{VVV}\\
&+12H^2_{VV}(H_{VW}^2-H_{VV}H_{WW}),\\
\ \\
S H_{VVVW} =&  H_{VVV}(2 H_W^2 H_{VVW}
  - 4 H_V H_W  H_{VWW} -  H_V^2 H_{WWW})+ 3 H_V^2 H_{VVW} H_{VWW}    \\
  &     - 3 H_W  H_{VV} H_{VW} H_{VVW} - 4 H_W  H_{VW}^2H_{VVV} - 6 H_V  H_{VW}^2 H_{VVW} \\
 &  + \frac{15}{2} H_W H_{VV}^2 H_{VWW} -
   6 H_V H_{VV} H_{VW} H_{VWW} +
   \frac{3}{2} H_V H_{VV}^2 H_{WWW} \\
&   - \frac{1}{2}H_W  H_{VV} H_{WW} H_{VVV}+
   \frac{3}{2} H_V H_{VV} H_{WW}  H_{VVW}+ 9 H_V  H_{VW} H_{WW} H_{VVV} \\
&+12H_{VV}H_{VW}(H_{VW}^2-H_{VV}H_{WW}),\\
\ \\
S H_{VVWW} =& 2(H_VH_{VWW}-H_WH_{VVW})^2 +2H_VH_W(H_{VVW}H_{VWW}-H_{VVV}H_{WWW})\\
&     - 8  H_{VW}^2(H_WH_{VVW}+H_VH_{VWW})    + 6 H_{W} H_{VV} H_{VW} H_{WWV} + 6 H_{V}  H_{WW} H_{VW}H_{VVW}\\
&   -  H_{VV} H_{WW}(H_{W} H_{VVW}  + H_{V}  H_{VWW} ) +   3 H_{W} H_{VV}^2 H_{WWW}  + 3 H_{V} H_{WW}^2H_{VVV}   \\
&+4(2H^2_{VW}+H_{VV}H_{WW})(H_{VW}^2-H_{VV}H_{WW}),\\
\ \\
S H_{VWWW} =& H_{WWW}(2 H_V^2  H_{VWW}
  - 4 H_V H_W  H_{VVW} -  H_W^2 H_{VVV})+ 3 H_W^2 H_{VVW} H_{VWW}    \\
  &     - 3 H_V  H_{WW} H_{VW} H_{VWW} - 4 H_V  H_{VW}^2H_{WWW} - 6 H_W  H_{VW}^2 H_{VWW} \\
 &  + \frac{15}{2} H_V H_{WW}^2 H_{VVW} -
   6 H_W H_{WW} H_{VW} H_{VVW} +
   \frac{3}{2} H_W H_{WW}^2 H_{VVV} \\
&   - \frac{1}{2}H_V  H_{VV} H_{WW} H_{WWW}+
   \frac{3}{2} H_W H_{VV} H_{WW}  H_{VWW}+ 9 H_W  H_{VW} H_{VV} H_{WWW} \\
&+12H_{WW}H_{VW}(H_{VW}^2-H_{VV}H_{WW}),\\
\ \\
S H_{WWWW} =&   2 (H_{V} H_{WWW} -   H_W  H_{VWW})^2 + 4 H_W^2 (H_{VWW}^2- H_{WWW} H_{VVW}) \\
&  +12 H_V H_{WW}^2H_{VWW}  - 12 H_V  H_{WW} H_{VW}H_{WWW} -
     24 H_W  H_{WW} H_{VW} H_{VWW}\\
     &+ 8 H_W  H_{VW}^2 H_{WWW}+  6 H_W H_{WW}^2 H_{VVW} + 10 H_W  H_{WW} H_{VV} H_{WWW}\\
&+12H^2_{WW}(H_{VW}^2-H_{VV}H_{WW}),
\ \\
\end{aligned}
\end{gather*}
where
$$
S=H_{W}^2H_{VV}-2H_{V}H_{W}H_{VW}+H_{V}^2H_{WW}.
$$
This system is manifestly symmetric under the interchange of $V$ and $W$, and can be represented in compact form as
\begin{equation}
\begin{array}{c}
Sd^4H=2d^3H dS-6(dH)^2\det (dN)+6dHd^2Hd(\det N)+\\
\ \\
12(H_VdH_W-H_WdH_V)(d^2H_VdH_W-d^2H_WdH_V)-
12\det N(d^2H)^2,
\end{array}
\label{HVW}
\end{equation}
where    $N$ is the Hessian matrix of $H$.
 We have verified that the system (\ref{HVW}) is in involution, and is invariant under a $10$-dimensional group of Lie-point symmetries generated by projective transformations of $V$ and $W$ along with  affine transformations of $H$,
$$
V\to \frac{aV+bW+c}{pV+qW+r}, ~~~ W\to \frac{\alpha V+\beta W+\gamma}{pV+qW+r}, ~~~ H\to \mu H+\nu.
$$
The corresponding infinitesimal generators are:
\begin{equation}
\begin{array}{c}
\noindent {\rm 2~translations}: ~~
\displaystyle  \frac{\partial}{\partial V}, ~~~ \frac{\partial}{\partial W};\\
\ \\
 \noindent {\rm 4~ linear~transformations}:
\displaystyle ~~~ V \frac{\partial}{\partial V}, ~~~ W\frac{\partial}{\partial V},  ~~~ V\frac{\partial}{\partial W}, ~~~ W\frac{\partial}{\partial W}; \\
\ \\
\noindent {\rm 2~projective~transformations}:
~~~
\displaystyle V^2\frac{\partial}{\partial V}+VW\frac{\partial}{\partial W}, ~~~
VW\frac{\partial}{\partial V}+W^2\frac{\partial}{\partial W};\\
\ \\
\noindent {\rm 2~affine~transformations~of} ~ H:
~~~
\displaystyle \frac{\partial}{\partial H}, ~~~ H\frac{\partial}{\partial H}.
 \end{array}
 \label{sym2}
 \end{equation}
 Projective transformations (\ref{sym2}) constitute the group of  `canonical' transformations of  equations (\ref{EP}): combined with appropriate linear changes of the independent variables $x, y, t$, they leave equations (\ref{EP}) form-invariant. Let us consider, for instance, the projective transformation $\tilde V=1/V, \ \tilde W=W/V$. A direct calculation shows that the transformed equations take the form
 $$
(\tilde VH_{\tilde V}+H)_t= (H_{\tilde V})_x+(\tilde WH_{\tilde V})_y, ~~~ (\tilde VH_{\tilde W})_t=(H_{\tilde W})_x+(\tilde WH_{\tilde W}+H)_y.
 $$
 They assume the original form (\ref{EP}) on the identification $t\to x, \ x \to t, \ y\to -y$. In other words, the projective invariance of the integrability conditions is a manifestation of the invariance of the Hamiltonian formalism (\ref{Hb3}) under arbitrary linear transformations of the independent variables. This is analogous to the well-known invariance of Hamiltonian structures of hydrodynamic type in $1+1$ dimensions under linear transformations of $x$ and $t$ \cite{form}. We emphasise that this invariance is {\it not }present in the case of constant coefficient Poisson brackets.

 \subsection{Classification of integrable potentials}

The analysis of this section is somewhat similar to the classification of integrable Lagrangians of the form $\int u_tg(u_x, u_y) \ dxdydt$ proposed in \cite{FerOd}. This suggests that there may be a closer link between these two classes of equations. The main result of this section is the following

 \medskip

 \noindent {\bf Theorem 3} {\it The generic  integrable potential  of type (\ref{EP})  is given by the series
$$
H=\frac{1}{Wg(V)}\left(1+\frac{1}{g_1(V)W^6}+\frac{1}{g_2(V)W^{12}}+... \right).
$$
Here $g(V)$ satisfies the fourth order ODE,
$$
g''''(2gg'^2-g^2g'')+2g^2g'''^2-20gg'g''g'''+16g'^3g'''+18gg''^3-18g'^2g''^2=0,
$$
whose general solution can be represented in parametric form as
$$
 g= w_2(t), ~~~ V=\frac{w_1(t)}{w_2(t)},
$$
where $w_1(t)$ and $w_2(t)$ are two linearly independent solutions of the hypergeometric equation  $t(1-t){d^2w}/{dt^2}-\frac{2}{9}w=0$. The coefficients $g_i(V)$ are certain explicit expressions in terms of $g(V)$, e.g., $g_1(V)=gg''-2g'^2$ and so on.
Degenerations of this solution correspond to
$$
H =  \frac{1}{Wg(V)}, ~~~
H = \frac{1}{WV},
~~~
H = V-W\log W,
~~~
H = V-W^2/2.
$$
}

\noindent In the  rest of this section we provide  details of the classification, and discuss various properties and representations of the generic solution.
The system (\ref{HVW}) for $H(V, W)$ is not straightforward to solve explicitly. We will start with the investigation of special solutions which are invariant under various one-parameter subgroups of the equivalence group. In this case the system of integrability conditions  for  $H(V, W)$ reduces to ODEs which are easier to solve.
 Up to  conjugation and normalisation, there exist four essentially different one-parameter subgroups of the projective group $SL(3)$, with the infinitesimal generators
 $$
 \alpha V \frac{\partial}{\partial V}+W \frac{\partial}{\partial W}, ~~~
  V \frac{\partial}{\partial V}+\frac{\partial}{\partial W}, ~~~
  W \frac{\partial}{\partial V}+ \frac{\partial}{\partial W}, ~~~  \frac{\partial}{\partial V}.
  $$
Combined with the operators  ${\partial}/{\partial H}, \ H{\partial}/{\partial H}$ this leads to the following list of eleven essentially different
`ansatzes' governing invariant solutions (in what follows we do not consider degenerate solutions for which the expression $S=H_{W}^2H_{VV}-2H_{V}H_{W}H_{VW}+H_{V}^2H_{WW}$ equals zero):

\noindent {\bf Case 1. }  Solutions invariant under the operator $  {\partial}/{\partial V}+ \partial/\partial H$ are described by the ansatz $H=V+F(W)$. The integrability conditions imply $F''F''''-2F'''^2=0$. Modulo the equivalence transformations this leads to integrable potentials
$H=V-W^2/2$ and $H=V-W\log W$, which constitute the last two cases of Theorem 3. Applying to the first potential   transformations from the equivalence group we obtain integrable potentials of the form
$$
H(V, W)=\frac{Q(V, W)}{l^2(V, W)}
$$
where $Q$ and $l$ are  quadratic and linear forms, respectively (not necessarily homogeneous). The integrability implies that the line $l=0$  is tangential to the conic $Q=0$ on the $V, W$ plane. Any such potential can be reduced to the form $H=V-W^2/2$ by a projective transformation which sends $l$ to the line at infinity.

\noindent {\bf Case 2. }  Solutions invariant under the operator $  {\partial}/{\partial W}+H \partial/\partial H$ are described by the ansatz $H=e^{ W}F(V)$. This case gives no non-trivial solutions.

\noindent {\bf Case 3. }  Solutions invariant under the operator $ W {\partial}/{\partial V}+ {\partial}/{\partial W}$ are described by the ansatz $H=F(V-W^2/2)$. A simple analysis leads to the only polynomial potential $H=V-W^2/2$, the same as in Case 1.

\noindent {\bf Case 4. }  Solutions invariant under the operator $ W {\partial}/{\partial V}+ {\partial}/{\partial W}+\partial/\partial H$ are described by the ansatz $H= W+F(V-W^2/2)$. In this case $F$  turns out to be linear so that, modulo translations in $W$, we again arrive at the same potential as in Case 1.

\noindent {\bf Case 5. }  Solutions invariant under the operator $ W {\partial}/{\partial V}+ {\partial}/{\partial W}+H \partial/\partial H$ are described by the ansatz $H=e^{ W}F(V-W^2/2)$. A detailed analysis shows that this case gives no non-trivial solutions.

\noindent {\bf Case 6. }  Solutions invariant under the operator $ V {\partial}/{\partial V}+ {\partial}/{\partial W}$ are described by the ansatz $H=F(W-\ln V)$. This case gives no non-trivial solutions.

\noindent {\bf Case 7. }  Solutions invariant under the operator $V {\partial}/{\partial V}+ {\partial}/{\partial W}+\partial/\partial H$ are described by the ansatz $H= W+F(W-\ln V)$. This case gives no non-trivial solutions.

\noindent {\bf Case 8. }  Solutions invariant under the operator $ V {\partial}/{\partial V}+ {\partial}/{\partial W}+\mu H \partial/\partial H$ are described by the ansatz $H=e^{\mu W}F(W-\ln V)$, also no non-trivial solutions.

\noindent {\bf Case 9. }  Solutions invariant under the operator $ \alpha V {\partial}/{\partial V}+ W{\partial}/{\partial W}$ are described by the ansatz $H=F(W^{\alpha}/ V)$. Here one can assume $\alpha \ne 0, 1$. A straightforward substitution implies that, without any loss of generality, one can assume  $F$ to be linear,  while the parameter $a$ can only take  two values: $a=2$ or $a=-1$. This results in the two rational potentials $H=W^2/V$ and $H=1/VW$.
Notice that they are related by the projective transformation $W\to 1/W, \ V\to V/W$.

\noindent {\bf Case 10. }  Solutions invariant under the operator $\alpha V {\partial}/{\partial V}+W {\partial}/{\partial W}+\partial/\partial H$ are described by the ansatz $H=\ln  W+F(W^{\alpha}/V)$. This case gives no non-trivial solutions.

\noindent {\bf Case 11. }  Solutions invariant under the operator $ \alpha V {\partial}/{\partial V}+ W{\partial}/{\partial W}+\mu H \partial/\partial H$ are described by the ansatz $H=W^{\mu}F(W^{\alpha}/V)$. A detailed  analysis shows that, modulo the equivalence group and the solutions already discussed  above, the only essentially new possibility corresponds to the  ansatz
 $H=\frac{1}{Wg(V)}$ (here  $\mu=-1, \alpha =0$). It leads to the fourth order ODE for $g= g(z)$,
\begin{equation}
g''''(2gg'^2-g^2g'')+2g^2g'''^2-20gg'g''g'''+16g'^3g'''+18gg''^3-18g'^2g''^2=0,
\label{g}
\end{equation}
which possesses a remarkable $SL(2,
R)$-invariance inherited from (\ref{sym2}):
\begin{equation}
\tilde z=\frac{\alpha z+ \beta }{\gamma z+ \delta}, ~~~ \tilde g= (\gamma z+ \delta) g;
\label{gsym}
\end{equation}
here $\alpha, \beta, \gamma, \delta$ are arbitrary constants such that $\alpha \delta -\beta \gamma=1$. Moreover, there is an obvious scaling symmetry $g\to \lambda g$. The equation (\ref{g}) can be linearised as follows. Introducing $h=g'/g$, which means factoring out the scaling symmetry, we first rewrite it in  the form
\begin{equation}
h'''(h'-h^2)-2h''^2+12hh'h''-4h^3h''-15h'^3+9h^2h'^2-3h^4h'+h^6=0;
\label{h}
\end{equation}
the corresponding symmetry group modifies to
\begin{equation}
\tilde z=\frac{\alpha z+ \beta }{\gamma z+ \delta}, ~~~ \tilde h= (\gamma z+ \delta)^2 h+\gamma(\gamma z+ \delta).
\label{hsym}
\end{equation}
We point out that the same symmetry  occurs in the case of the Chazy
equation, see \cite{A1}, p. 342. The presence of the $SL(2, R)$-symmetry of
this type implies the linearisability of the equation under study.
One can formulate the following general statement which is, in fact,
contained in  \cite{Clarkson}.

\medskip

\noindent {\bf Proposition 1.}  {\it Any  third order ODE of the form $F(z, h, h',
h'', h''')=0$, which is invariant under the action of $SL(2, R)$ as
specified by (\ref{hsym}), can be linearised by a substitution
\begin{equation}
 h=\frac{d}{dz}\ln w_2, ~~~ z=\frac{w_1}{w_2}
\label{sub}
\end{equation}
where $w_1(t)$ and $w_2(t)$ are two linearly independent solutions of a linear equation $d^2w/dt^2=V(t)  w$ with the Wronskian $W$ normalised as  $W=w_2dw_1/dt-w_1dw_2/dt=1$. The potential $V(t)$ depends on the given third order ODE, and can be efficiently reconstructed.

In particular, the general solution of the equation (\ref{h}) is given by parametric formulae (\ref{sub})
where $w_1(t)$ and $w_2(t)$ are two linearly independent solutions of the hypergeometric equation $d^2w/dt^2=\frac{2}{9}\frac{1}{t(1-t)} w$ with $W=1$.}

\medskip

\centerline {\bf Proof:}

\medskip

\noindent  Our presentation  follows \cite{FerOd}. Let us consider a linear ODE $d^2w/dt^2=V(t)  w$,
 take two linearly independent solutions
$w_1(t)$, $w_2(t)$ with the Wronskian $W=1$, and introduce  new dependent and independent variables  $h, z$ by parametric relations
$$
 h=\frac{d}{dz}\ln w_2, ~~~ z=\frac{w_1}{w_2}.
$$
Using the  formulae $dt/dz=w_2^2$ and $ h=w_2dw_2/dt$, one obtains the identities
$$
h'-h^2=w_2^4\  V,
$$
$$
h''-6hh'+4h^3=w_2^6\ dV/dt,
$$
$$
h'''-12hh''-6(h')^2+48h^2h'-24h^4=w_2^8\ d^2V/dt^2,
$$
where prime denotes differentiation with respect to $z$.  Thus, one arrives at the relations
$$
\begin{array}{c}
I_1=\displaystyle{\frac{(h''-6hh'+4h^3)^2}{(h'-h^2)^3}=\frac{(dV/dt)^2}{V^3}},   \\
\ \\
I_2=\displaystyle{\frac{h'''-12hh''-6(h')^2+48h^2h'-24h^4}{(h'-h^2)^2}=\frac{d^2V/dt^2}{V^2}}. \\
\end{array}
$$
We point out that $I_1$ and $I_2$ are the simplest second and third order differential invariants of the action (\ref{hsym}) whose infinitesimal generators, prolonged to the third order jets $z, h, h', h'', h'''$, are of the form
$$
\begin{array}{c}
X_1=\partial_z, ~~~ X_2=z\partial_z-h\partial_h-2h'\partial_{h'}-3h''\partial_{h''}-4h'''\partial_{h'''}, \\
\ \\
X_3=z^2\partial_z-(2zh+1)\partial_h-(2h+4zh')\partial_{h'}-(6h'+6zh'')\partial_{h''}-(12h''+8zh''')\partial_{h'''};
\end{array}
$$
notice the standard commutation relations $[X_1,  X_2]=X_1, \  [X_1,
X_3]=2X_2, \ [X_2,  X_3]=X_3$. One can verify that the Lie
derivatives of $I_1, I_2$ with respect to $X_1, X_2, X_3$ are indeed
zero. Thus, any third order ODE which is invariant under the $SL(2,
R)$-action (\ref{hsym}), can be represented in the form
$I_2=F(I_1)$ where $F$ is an arbitrary function of one variable. The
corresponding potential $V(t)$ has to satisfy the equation
$\displaystyle{\frac{d^2V/dt^2}{V^2}=F\left(\frac{(dV/dt)^2}{V^3} \right)}.$

This simple scheme produces many the well-known equations, for instance,  the relation $I_2=-24$  implies the Chazy equation for $h$, that is,
$h'''-12hh''+18(h')^2=0$. The corresponding potential satisfies the equation $d^2V/dt^2=-24V^2$.

Similarly, the choice $I_2=I_1-8$ results in the ODE $h'''=4hh''-2(h')^2+\displaystyle{\frac{(h''-2hh')^2}{h'-h^2}}$ which, under the substitution $h=y/2$, coincides with the equation (4.7) from \cite{A2}.
The potential $V$ satisfies the equation $Vd^2V/dt^2=(dV/dt)^2-8V^3$.

\medskip

The relation $I_2=I_1-9$ gives $h'''(h'-h^2)=h^6-3h^4h'+9h^2(h')^2-3(h')^3-4h^3h''+(h'')^2$. This equation appeared in \cite{FerOd} in the context of first order integrable Lagrangians. The corresponding potential $V$ satisfies the equation $Vd^2V/dt^2=(dV/dt)^2-9V^3$.

\medskip

Finally, the relation $I_2=2I_1+9$ coincides with the equation (\ref{h}). The corresponding potential $V$ satisfies the equation $Vd^2V/dt^2=2(dV/dt)^2+9V^3$. It remains to point out that the general solution to the last  equation for $V$ is given by $V=-\frac{2}{9}\frac{1}{t^2+at+b}$. Without any loss of generality one can set $V=\frac{2}{9}\frac{1}{t(1-t)}$. It this case the linear equation $d^2w/dt^2=V(t)  w$ takes the  hypergeometric form corresponding to the parameter values $a=-\frac{1}{3}, \ b=-\frac{2}{3},\  c=0$: $t(1-t){d^2w}/{dt^2}-\frac{2}{9}w=0$. This finishes the proof of Proposition 1.


\medskip

As  $h=g'/g$, this immediately implies the following formula for the general solution of (\ref{g}):

\noindent{\bf Proposition 2.} {\it The general solution of the equation (\ref{g}) is given by parametric formulae
$$
 g= w_2, ~~~ z=\frac{w_1}{w_2},
$$
where $w_1$ and $w_2$ are two linearly independent solutions to the hypergeometric equation $t(1-t){d^2w}/{dt^2}-\frac{2}{9}w=0$.}

\medskip

\noindent {\bf Remark.} Expansions at zero give
\begin{equation*}
w_1 = w_2\ln t + \frac{9}{2}+\frac{t^2}{3}+\frac{211 t^3}{1458}+...,
 ~~~ w_2 = t+\frac{t^2}{9}+\frac{10 t^3}{243}+...,
\end{equation*}
so that
\begin{equation*}
g=w_2 = t+\frac{t^2}{9}+\frac{10 t^3}{243}+..., ~~~~ z= \frac{w_1}{w_2}=\ln t + \frac{9}{2 t}-\frac{1}{2}+\frac{11 t}{54}+\frac{34 t^2}{729}+\frac{715 t^3}{39366}+....
\end{equation*}
Solving the first relation for $t$ in terms of $g$ and substituting into the second, one gets an implicit relation connecting $g$ and $z$,
\begin{equation*}
z = \ln g  - \frac{9}{g} + \frac{1}{2} + \frac{112 g}{27} + \frac{289 g^2}{486} + \frac{4381 g^3}{39366} +....
\end{equation*}

\medskip

\noindent Similarly, expansions at infinity give

$$
{w_1} = {t^{2/3}}\left( 1-\frac{1}{3t}-\frac{2}{45t^2} -...\right), ~~~ {w_2} = {t^{1/3}}\left(1-\frac{1}{6t}-\frac{5}{126 t^2} -...\right),
$$
$$
z = \frac{w_1}{w_2}= {t^{1/3}}\left(1-\frac{1}{6t}-\frac{41}{1260t^2}-... \right),
$$
 so that $g = w_2$ can be represented explicitly as
$$
g(z)=z \left(1-\frac{a}{z^6}-\frac{1165}{143} \frac{ a^2}{ z^{12}}-\frac{5280035 }{46189}\frac{a^3}{ z^{18}}-...\right),
$$
 here $a=1/140$. For other values of $a$ this formula represents the general  solution to (\ref{g}) in the form $g=z(1-a/z^6-b/z^{12}-c/z^{18}-...)$. The corresponding potential $H(V, W)$ takes the form
 $$
 H(V, W)=\frac{1}{Wg(V)}=\frac{1}{VW}\left(1+\frac{a}{V^6}+\frac{1308}{143}\frac{a^2}{V^{12}}+...   \right).
 $$
It can be viewed as a perturbation of the potential $H=\frac{1}{VW}$ constructed before. Computer experiments show that the `generic'
integrable potential $H(V, W)$ can be represented as
\begin{eqnarray*}
H=\sum_{j, k\geq 0} \frac{a_{jk}}{V^{6j+1}W^{6k+1}}= \frac{1}{V W}\left (1 +\frac{1}{V^6} +\frac{1}{W^6} -\frac{24}{V^6 W^6} +\frac{1308}{143 V^{12}} +\frac{1308}{143 W^{12}}
 + \dots \right).
\end{eqnarray*}
An alternative representation of this solution,
$$
H=\frac{1}{Wg(V)}\left(1+\frac{1}{g_1(V)W^6}+\frac{1}{g_2(V)W^{12}}+... \right),
$$
 can be obtained be rearranging terms in the above sum. The substitution of this expression into the integrability conditions implies that $g(V)$ has to satisfy the fourth order ODE (\ref{g}),  $g_1(V)$ is expressed in terms of $g(V)$ via the Rankin-Cohen-type operation, $g_1(V)=gg''-2g'^2$ (recall that, due to (\ref{gsym}), $g(V)$ transforms as a modular form of weight 1), and so on.   This finishes the proof of Theorem 3.

 In Sect. 5 we provide a parametrisation of the generic integrable potential  by generalised hypergeometric functions.

\subsection {Dispersionless Lax pairs}

Here we present Lax pairs for some of the simplest potentials found in the previous section.

\noindent {\bf Case $H=V-W^2/2$.} The corresponding system (\ref{EP}) takes the form
$$
W_y=WW_x-2V_x, ~~~ W_t=VW_x-5WV_x+3W^2W_x-V_y,
$$
and possesses the Lax pair
$$S_y = -  WS_x+\frac{1}{S_x^2},  ~~~
S_t =   (V-W^2)S_x+\frac{W}{S_x^2}-\frac{2}{5S_x^5}.
$$

\noindent {\bf Case $H=V-W\log W$.} The corresponding system (\ref{EP}) takes the form
$$
W_y=(W\ln W-2V)_x, ~~~ W_t /W=(V\ln W+V)_x+(2W\ln W+W-V)_y,
$$
and possesses the Lax pair
$$
S_y = -S_x(\ln W +1)+ p(S_x), ~~~
S_t = S_x(V -W\ln W-W)-\frac{1}{2}S_xp'(S_x)W
$$
where $p(z)$ satisfies   the ODE
$zp'' +p'^2+3p' = 0.$ This gives $p'=\frac{3}{z^3-1}$ so that
\be
p(S_x)=  \ln(S_x-1) +\epsilon \ln(S_x-\epsilon) + \epsilon^2 \ln(S_x-\epsilon^2), ~~~ \epsilon=e^{\frac{2\pi i}{3}}.
\ee

\noindent {\bf Case $H=\frac{1}{WV}$.}  The corresponding system (\ref{EP}) takes the form
\begin{equation}
\left(\frac{1}{V^2W}\right)_t=\left(\frac{1}{V^2}\right)_y, ~~~ \left(\frac{1}{VW^2}\right)_t=\left(\frac{1}{W^2}\right)_x,
\label{Lag}
\end{equation}
and possesses the Lax pair
$$
S_x = a(S_t)/V,  ~~~ S_y=  b(S_t)/W
$$
where the functions $a(z)$ and $b(z)$, $z=S_t$, satisfy a pair of ODEs
$$
a'=1-\frac{a}{2b}, ~~~ b'=1-\frac{b}{2a}.
$$
These equations can be solved in parametric form as
$$
a(z)=\frac{\wp'(p)+\lambda}{2\wp(p)}, ~~~ b(z)=\frac{\wp'(p)-\lambda}{2\wp(p)}, ~~~ z=-2\zeta(p),
$$
where $\wp(p)$ and $\zeta(p)$ are the Weierstrass functions, $\wp'^2=4\wp^3+\lambda^2, \ \zeta'=-\wp$. The resulting Lax pair can be written in parametric form as
$$
S_x=\frac{1}{V} \frac{\wp'(p)+\lambda}{2\wp(p)}, ~~~ S_y=\frac{1}{W} \frac{\wp'(p)-\lambda}{2\wp(p)}, ~~~ S_t=-2\zeta(p),
$$
or, equivalently,
$$
S_xS_y(VS_x-WS_y)=\frac{\lambda}{VW}, ~~~ S_xS_y=\frac{\wp(p)}{VW}, ~~~ S_t=-2\zeta(p).
$$
Notice that  equations (\ref{Lag})  coincide with the Euler-Lagrange equations corresponding to the Lagrangian density $\int \frac{u_xu_y}{u_t^2} dxdydt$  upon setting $V=\frac{u_t}{2u_x}, \ W=\frac{u_t}{2u_y}$ (we thank Maxim Pavlov for pointing this out). In this context, the above Lax pair  appeared previously in \cite{max1}.


\section{Godunov systems and generalized hypergeometric functions}

In this section we develop the general theory of  Godunov's systems \cite{Godunov}, and describe the Godunov form of $n$-component quasilinear systems constructed in \cite{Odesskii0, Odesskii2} in terms of generalized hypergeometric functions. In particular, this provides a Godunov representation for any generic 2-component integrable system \cite{Fer2, Odesskii0}. Applied to the system (\ref{EP}), this  construction gives a parametrisation of the generic integrable potential $H(V, W)$ in the form (\ref{HypO}).

In Sect. 5.1 we recall the main aspects of the Godunov representation, and clarify its symmetry properties. In Sect. 5.2 we construct the  Godunov form for quasilinear systems found in \cite{Odesskii0, Odesskii2} in terms of generalised hypergeometric functions. In Sect. 5.3 we specialise this construction to  integrable potentials $H(V, W)$ of the type (\ref{EP}), thus proving Theorem 4.

\subsection{Godunov systems}

A $2+1$ dimensional quasilinear system in $n$ unknowns ${\bf v}=(v^1, ..., v^n)$  is said to possess a  Godunov representation \cite{Godunov}  if it can be written in the conservative form,
\begin{equation}
(F_{0,i})_t+(F_{1,i})_x+(F_{2,i})_y=0, ~~~~~ i=1,\dots, n,
\label{Godunov}
\end{equation}
where potentials $F_0, F_1, F_2$ are functions of ${\bf v}$, and  $F_{\alpha,j}=\partial F_{\alpha}/\partial v^j, \ \alpha=0, 1, 2$. Any such system automatically possesses an extra conservation law,
\begin{equation}
{\cal L}(F_0)_t+{\cal L}(F_1)_x+{\cal L}(F_2)_y=0,
\label{Legendre}
\end{equation}
where ${\cal L}$ denotes the Legendre transform: ${\cal L}(F_{\alpha})=F_{\alpha,k}v^k-F_{\alpha}$. Many systems of physical origin are known to be representable in the Godunov form. This representation is widely used for analytical/numerical treatment of quasilinear systems.

Note that the systems (\ref{vw})-(\ref{EP}) are written in the Godunov form. For example, in the case (\ref{EP}) we have $n=2,~v^1=V,~v^2=W,~F_0=-H,~F_1=VH,~F_2=WH$.  Recall that a $2+1$ dimensional quasilinear system of $n$ equations for $n$ unknowns possesses a Godunov form iff it possesses $n+1$ conservation laws of hydrodynamic type. The following fact will be useful:

\medskip

\noindent {\bf Proposition 1.} {\it The Godunov representation (\ref{Godunov}) is form-invariant under the projective action of  $GL_{n+1}$ defined as
$$
\tilde v^i=\frac{l^i({\bf v})}{l({\bf v})},  ~~~ \tilde F_{\alpha}=\frac{F_{\alpha}}{l({\bf v})},
$$
here $l^i, \ l$ are linear (inhomogeneous) forms in ${\bf v}$.}

\medskip

\centerline{\bf Proof:}

\medskip

\noindent With any Godunov system we associate the following geometric objects. Let us consider an auxiliary  $(n+1)$-dimensional affine space $A^{n+1}$ with coordinates $x_1, \dots,  x_n,  x_{n+1}$. With any potential $F_{\alpha}$ we associate an  $n$-parameter family of hyperplanes,
\begin{equation}
x_{n+1}-x_kv^k+F_{\alpha}({\bf v})=0.
\label{planes}
\end{equation}
The envelope of this family is a hypersurface $M^n_{F_{\alpha}}\subset A^{n+1}$ defined parametrically as
$$
(x_1, \dots, x_n, \ x_{n+1})=(F_{\alpha,1}, \dots, F_{\alpha,n}, \ {\cal L} (F_{\alpha})).
$$
Notice that components of its position vector are the conserved densities appearing in Eqs. (\ref{Godunov}), (\ref{Legendre}). By construction, hypersurfaces $M^n_{F_0}, \ M^n_{F_1}$ and $ M^n_{F_2}$ have parallel tangent hyperplanes at the points corresponding to the same values of the parameters $v^i$.  Let us now apply an arbitrary affine transformation in $A^{n+1}$, $\tilde x=Ax$, where $A$ is a constant $(n+1)\times (n+1)$ matrix. This will transform
Eq. (\ref{planes}) to
\begin{equation}
\tilde x_{n+1}-\tilde x_k\tilde v^k+\tilde F_{\alpha}(\tilde {\bf v})=0
\label{planes1}
\end{equation}
where the transformation ${\bf v} \to \tilde {\bf v}$ will automatically be projective:
$$
\tilde v^i=\frac{l^i({\bf v})}{l({\bf v})},  ~~~ \tilde F_{\alpha}=\frac{F_{\alpha}}{l({\bf v})};
$$
here $l^i, \ l$ are linear (inhomogeneous) forms in ${\bf v}$. Since affine transformations preserve the properties of being parallel/tangential, they naturally act on the class of Godunov's systems.

\subsection{Godunov form of integrable quasilinear systems and generalized hypergeometric functions}

Let $H_{s_1,...,s_{n+2}}$ be the space of solutions of the system
\begin{equation} \label{hyper}
\frac{\partial^2 h}{\partial u_i \partial
u_j}=\frac{s_i}{u_i-u_j}\cdot \frac{\partial h}{\partial
u_j}+\frac{s_j}{u_j-u_i}\cdot \frac{\partial h}{\partial
u_i},~~~i,j=1,...,n, \qquad i\ne j,
\end{equation}
and
\begin{equation} \begin{array}{c} \label{hyper1}
\displaystyle \frac{\partial^2 h}{\partial u_i \partial
u_i}=-\Big(1+\sum_{j=1}^{n+2} s_j\Big) \frac{s_i}{u_i (u_i-1) }\, h+
\frac{s_i}{u_i (u_i-1)} \sum_{j\ne i}^n \frac{u_j
(u_j-1)}{u_j-u_i}\cdot
\frac{\partial h}{\partial u_j}+\\[7mm]
\displaystyle \Big(\sum_{j\ne i}^n \frac{s_j}{u_i-u_j}+
\frac{s_i+s_{n+1}}{u_i}+ \frac{s_i+s_{n+2}}{u_i-1}\Big)
\frac{\partial h}{\partial u_i},
\end{array}
\end{equation}
for one unknown function $h(u_1, \dots, u_n)$. Here $s_1,...,s_{n+2}$ are arbitrary constants. Elements of $H_{s_1,...,s_{n+2}}$ are examples of the so-called generalised hypergeometric functions, see \cite{gel}, \cite{Odesskii2}. It is known that $\dim H_{s_1,...,s_{n+2}}=n+1$. Let $g_0,g_1,...,g_n$ be a basis of $H_{s_1,...,s_{n+2}}$. Choose a time $t_q$ for each element $g_q$ of this basis. Here $q$ runs from 0
to $n$. It is known \cite{Odesskii0}, \cite{Odesskii2} that for each pairwise distinct $q, r, s$ running from 0 to $n$ the system
$$\sum_{1\leq i\leq n,i\ne
j}(g_{q,u_j}g_{r,u_i}-g_{r,u_j}g_{q,u_i})\frac{u_j(u_j-1)u_{i,t_{s}}-u_i(u_i-1)u_{j,t_{s}}}{u_j-u_i}+
\sigma\cdot (g_{q}g_{r,u_j}-g_{r}g_{q,u_j})u_{j,t_{s}}+$$
\begin{equation}
\label{3dsyst} \sum_{1\leq i\leq n,i\ne
j}(g_{r,u_j}g_{s,u_i}-g_{s,u_j}g_{r,u_i})\frac{u_j(u_j-1)u_{i,t_{q}}-u_i(u_i-1)u_{j,t_{q}}}{u_j-u_i}+
\sigma\cdot (g_{r}g_{s,u_j}-g_{s}g_{r,u_j})u_{j,t_{q}}+
\end{equation}
$$\sum_{1\leq i\leq n,i\ne
j}(g_{s,u_j}g_{q,u_i}-g_{q,u_j}g_{s,u_i})\frac{u_j(u_j-1)u_{i,t_{r}}-u_i(u_i-1)u_{j,t_{r}}}{u_j-u_i}+
\sigma\cdot(g_{s}g_{q,u_j}-g_{q}g_{s,u_j})u_{j,t_{r}}=0,$$ where
$j=1,...,n,\,$ $\sigma=1+s_1+...+s_{n+2}$,  possesses a disperssionless Lax representation and an infinity of hydrodynamic reductions. Moreover, any generic integrable 3-dimensional hydrodynamic type system with two unknowns is isomorphic to a system of the form (\ref{3dsyst}), see \cite{Odesskii0}. It is also known that the system (\ref{3dsyst}) possesses $n+1$ conservation laws of hydrodynamic type, see \cite{Fer2} for $n=2$ and \cite{Odesskii2} for general $n$. Therefore, it possesses a Godunov representation which can be constructed explicitly in the following way.

\begin{theorem} {\it Let $h_0,h_1,...,h_n$ be a basis of $H_{2s_1,...,2s_{n+2}}$. For each $\alpha\ne \beta=0,1,...,n$ let $f_{\alpha,\beta}$ be a solution of the inhomogeneous linear system
\begin{equation} \label{nonhhyper}
\frac{\partial^2 h}{\partial u_i \partial
u_j}=\frac{2s_i}{u_i-u_j}\cdot \frac{\partial h}{\partial
u_j}+\frac{2s_j}{u_j-u_i}\cdot \frac{\partial h}{\partial
u_i}+\frac{\frac{\partial g_{\alpha}}{\partial u_i}\frac{\partial g_{\beta}}{\partial u_j}-\frac{\partial g_{\alpha}}{\partial u_j}\frac{\partial g_{\beta}}{\partial u_i}}{u_i-u_j},~~~i,j=1,...,n, \qquad i\ne j,
\end{equation}
and
\begin{equation} \begin{array}{c} \label{nonhhyper1}
\displaystyle \frac{\partial^2 h}{\partial u_i \partial
u_i}=-\Big(1+2\sum_{j=1}^{n+2} s_j\Big) \frac{2s_i}{u_i (u_i-1) }\, h+
\frac{2s_i}{u_i (u_i-1)} \sum_{j\ne i}^n \frac{u_j
(u_j-1)}{u_j-u_i}\cdot
\frac{\partial h}{\partial u_j}+\\[7mm]
\displaystyle \Big(\sum_{j\ne i}^n \frac{2s_j}{u_i-u_j}+
\frac{2s_i+2s_{n+1}}{u_i}+ \frac{2s_i+2s_{n+2}}{u_i-1}\Big)
\frac{\partial h}{\partial u_i}-
\end{array}
\end{equation}
$$\sum_{j\ne i}\frac{\frac{\partial g_{\alpha}}{\partial u_i}\frac{\partial g_{\beta}}{\partial u_j}-\frac{\partial g_{\alpha}}{\partial u_j}\frac{\partial g_{\beta}}{\partial u_i}}{u_i-u_j}\cdot\frac{u_j(u_j-1)}{u_i(u_i-1)}-(1+s_1+...+s_{n+2})\frac{\frac{\partial g_{\alpha}}{\partial u_i}g_{\beta}-\frac{\partial g_{\beta}}{\partial u_i} g_{\alpha}}{u_i(u_i-1)}.$$
 Define new coordinates $v_1,...,v_n$ and  the functions $F_{\alpha,\beta}(v_1,...,v_n)$ by
\begin{equation} \label{vF}
v_i=\frac{h_i(u_1,...,u_n)}{h_0(u_1,...,u_n)},~~~F_{\alpha,\beta}=\frac{f_{\alpha,\beta}}{h_0(u_1,...,u_n)}.
\end{equation}
Then, in coordinates $v_1,...,v_n$, the system (\ref{3dsyst})  takes the Godunov form
\begin{equation}
 \label{godhyper}
\left(\frac{\partial F_{r,s}}{\partial v_i}\right)_{t_q}+\left(\frac{\partial F_{s,q}}{\partial v_i}\right)_{t_r}+\left(\frac{\partial F_{q,r}}{\partial v_i}\right)_{t_s}=0, ~~~ i=1, \dots, n.
 \end{equation}}
\end{theorem}

\centerline {\bf Proof:}

\medskip

\noindent Substituting (\ref{vF}) into (\ref{godhyper}) and calculating  derivatives of $f_{\alpha,\beta}$ by virtue of (\ref{nonhhyper}), (\ref{nonhhyper1}) we obtain conservation laws of the system (\ref{3dsyst}) as found in \cite{Odesskii2}. More precisely, the left hand side of (\ref{godhyper}) is equal to
$$\sum_{j=1}^n\frac{(-1)^{i+j}\det W_{i,j}}{u_j(u_j-1)\det W}R_j$$
where $R_j$ is the left hand side of (\ref{3dsyst}), the matrix $W$ is defined as $W=(w_{\alpha,\beta})$ where $w_{0,\beta}=h_{\beta}(u_1,...,u_n),~w_{\alpha,\beta}=\frac{\partial h_{\beta}}{\partial u_{\alpha}}$ if $\alpha\ne0$ and $W_{i,j}$ is the $n\times n$-minor of $W$ obtained by eliminating  a column with $h_i$ and a row with $\frac{\partial}{\partial u_j}$. In other words, the left hand side of (\ref{godhyper}) is equal to
$$\sum_{j=1}^n\frac{(W^{-1})_{i,j}}{u_j(u_j-1)}R_j,$$
so that  (\ref{godhyper}) holds identically modulo (\ref{3dsyst}).

\medskip

\noindent {\bf Remark 1.} The system (\ref{godhyper}) possesses a natural action of the group $GL_{n+1}\times GL_{n+1}$. Namely, the first copy of $GL_{n+1}$ acts on the times $t_0,...,t_n$, and in the same way on the basis $g_0, ..., g_n$. This action corresponds to a change of  basis $g_0, ..., g_n$ in $H_{s_1, ..., s_{n+2}}$. The second copy of $GL_{n+1}$ acts according to the Proposition 1. This action corresponds to a change of  basis $h_0,...,h_n$ in $H_{2s_1, ..., 2s_{n+2}}$.

\medskip

\noindent {\bf Remark 2.} Note that $F_{\alpha,\beta}$ is defined up to an arbitrary linear combination of $1,v_1,...,v_n$ and, therefore, $f_{\alpha,\beta}$ is defined up to an arbitrary linear combination of $h_0,h_1,...,h_n$. This explains why $f_{\alpha,\beta}$ satisfies a linear non-homogeneous system with the same homogeneous part as for elements from $H_{2s_1,...,2s_{n+2}}$. Moreover, the structure of the non-homogeneous part of the system (\ref{nonhhyper}), (\ref{nonhhyper1}) containing linear combinations of $\frac{\partial g_{\alpha}}{\partial u_i}\frac{\partial g_{\beta}}{\partial u_j}-\frac{\partial g_{\alpha}}{\partial u_j}\frac{\partial g_{\beta}}{\partial u_i}$ and $\frac{\partial g_{\alpha}}{\partial u_i}g_{\beta}-\frac{\partial g_{\beta}}{\partial u_i} g_{\alpha}$ is dictated by the action of $GL_{n+1}$ on the times $t_0,...,t_n$, and in the same way on $g_0,...,g_n$, compare with (\ref{3dsyst}).

\noindent {\bf Remark 3.} It was proven in \cite{Odesskii0} that any 2-component integrable system is isomorphic to a member of the family (\ref{3dsyst}) with $n=2$, or its appropriate limit. Therefore,  Theorem 5 gives, in particular,  a description of the Godunov form for any generic 2-components integrable system. If $n>2$  there exist $n$-component integrable systems which do not belong to the family (\ref{3dsyst}) or its degenerations (see, for example, \cite{Odesskii3}). However,  $n$-component systems constructed in \cite{Odesskii3}) have $n$ conservation laws only and, therefore, do not possess a Godunov form. Probably, the only integrable systems possessing Godunov's form should belong to the family (\ref{3dsyst}) or its appropriate degenerations.

\subsection{Application to integrable potentials H(V, W)}

Let us now apply these results to the system (\ref{EP}). Set $n=2$, choose a triple of indices $s=0,~q=1,~r=2$ and set $t_0=-t,~t_1=x,~t_2=y$ where $t,~x,~y$ are the independent variables in (\ref{EP}). Note that the bases in $H_{s_1,...,s_{n+2}}$ and $H_{2s_1,...,2s_{n+2}}$ are no longer independent since we have constraints on the functions $F_{1,2},~F_{2,0},~F_{0,1}$, namely, $F_{2,0}=v_1F_{1,2},~F_{0,1}=v_2F_{1,2}$. To obtain (\ref{EP}) it remains to set $F_{1,2}=H, ~v_1=V,~v_2=W$. Therefore, $f_{1,2}=Hh_0,~f_{2,0}=Hh_1,~f_{0,1}=Hh_2.$ Moreover, we must have
$$
h_i=a(u_1,u_2)(g_jg_{k,u_1}-g_kg_{j,u_1})+b(u_1,u_2)(g_jg_{k,u_2}-g_kg_{j,u_2})+c(u_1,u_2)(g_{j,u_2}g_{k,u_1}-g_{j,u_1}g_{k,u_2})$$
where $i,j,k$ is a cyclic permutation of $0,1,2$. Indeed, $h_i$ must have the same structure  in $g_0,~g_1,~g_2$ as coefficients at $u_{j,t_i}$ in (\ref{3dsyst}).  Substituting the expressions for $h_0,h_1,h_2$ and  $f_{1,2},~f_{2,0},~f_{0,1}$ into the equations for $H_{2s_1,...,2s_{n+2}}$ and  (\ref{nonhhyper}), (\ref{nonhhyper1}), respectively, and using the equations (\ref{hyper}), (\ref{hyper1}) for $g_0,g_1,g_2$, we obtain that $s_1=s_2=-s_3=s_4=-\frac{1}{3}$, along with  the following expressions for the functions $h_0,~h_1,~h_2$:
$$h_0(u_1,u_2)=C\Big(\frac{u_2}{u_2-1}(g_1g_{2,u_1}-g_2g_{1,u_1})+$$
$$
\frac{u_1}{u_1-1}(g_1g_{2,u_2}-g_2g_{1,u_2})+{3(u_1-u_2)}(g_{1,u_2}g_{2,u_1}-g_{1,u_1}g_{2,u_2})\Big),$$
$$h_1(u_1,u_2)=C\Big(\frac{u_2}{u_2-1}(g_2g_{0,u_1}-g_0g_{2,u_1})+$$
$$\frac{u_1}{u_1-1}(g_2g_{0,u_2}-g_0g_{2,u_2})+{3(u_1-u_2)}(g_{2,u_2}g_{0,u_1}-g_{2,u_1}g_{0,u_2})\Big),$$
$$h_2(u_1,u_2)=C\Big(\frac{u_2}{u_2-1}(g_0g_{1,u_1}-g_1g_{0,u_1})+$$
$$\frac{u_1}{u_1-1}(g_0g_{1,u_2}-g_1g_{0,u_2})+{3(u_1-u_2)}(g_{0,u_2}g_{1,u_1}-g_{0,u_1}g_{1,u_2})\Big),$$
where $$C=(u_1-1)^{2/3}(u_2-1)^{2/3}(u_1-u_2)^{-1/3}.$$
We also obtain the following system for $H$,
$$H_{u_1}=\frac{u_2-1}{C (u_1-u_2)}, ~~ H_{u_2}=\frac{u_1-1}{C(u_2-u_1)}.$$
The solution of this system reads
$$H(u_1,u_2)=G\Big(\frac{u_1-1}{u_2-1}\Big)$$
where $G^{\prime}(t)=\frac{1}{t^{2/3}(t-1)^{2/3}}$.
Summarizing, we obtain the proof of the following

\medskip

\noindent {\bf Theorem 4} {\it Let the potential $H(V, W)$ be defined parametrically in the form
$$~~ H=G\Big(\frac{u_1-1}{u_2-1}\Big), ~~ V=\frac{h_1}{h_0},~~ W=\frac{h_2}{h_0},$$
where $G^{\prime}(t)=\frac{1}{t^{2/3}(t-1)^{2/3}}$ and $h_0,h_1,h_2$ are three linearly independent solution of the hypergeometric system
$$
u_1(1-u_1)h_{u_1,u_1} - \frac{4}{3}u_1h_{u_1}-\frac{2}{9}h= \frac{2}{3}\frac{u_1(u_1-1)}{u_1-u_2}h_{u_1} +\frac{2}{3}\frac{u_2 (u_2-1)}{u_2-u_1} h_{u_2},
$$
$$
u_2(1-u_2)h_{u_2,u_2} - \frac{4}{3}u_2h_{u_2}-\frac{2}{9}h= \frac{2}{3}\frac{u_1(u_1-1)}{u_1-u_2}h_{u_1} +\frac{2}{3}\frac{u_2 (u_2-1)}{u_2-u_1} h_{u_2},
$$
$$
h_{u_1,u_2} =  \frac{2}{3}\frac{h_{u_2}-h_{u_1}}{u_2-u_1};
$$
(this system coincides with (\ref{hyper}), (\ref{hyper1}) for the values of constants $s_1=s_2=-s_3=s_4=-\frac{2}{3}$). Then $H(V,W)$ provides the generic solution to the system (\ref{HVW}). In particular, it does not possess any continuous symmetry from the equivalence group.}

\section{Appendix. The method of hydrodynamic reductions. Derivation of the integrability conditions}

Applied to a $(2+1)$-dimensional system (\ref{2+1}), the method of hydrodynamic reductions  consists of seeking multi-phase solutions in the form
$$
{\bf u} (x, y, t)={\bf u}(R^1, ..., R^n)
$$
where the phases $R^i(x, y, t)$ are required to satisfy a pair of
 $(1+1)$-dimensional systems of hydrodynamic type,
$$
R^i_t = \lambda^i (R) R_x^i, ~~~ R^i_y = \mu^i (R) R_x^i.
$$
Solutions of this form,  known as `non-linear interactions of $n$ planar simple waves'  \cite{Sidorov, Bu, P}, have been extensively discussed in gas dynamics;  later, they reappeared in the context of the dispersionless KP hierarchy, see \cite{GibTsa96, GibTsa99} and references therein. Effectively, one decouples the $(2+1)$-dimensional system (\ref{2+1}) into a pair of commuting $n$-component $(1+1)$-dimensional systems, which are called its hydrodynamic reductions.
Substituting the ansatz ${\bf u}(R^1, ..., R^n)$ into (\ref{2+1})  one obtains
\begin{equation}
(\lambda^i I_n-A-\mu^iB)\ \partial_i{\bf u}=0, ~~~~~ i=1, ..., n,
\label{reduction}
\end{equation}
$\partial_i=\partial/\partial R^i$, implying that both characteristic speeds $\lambda^i$ and $\mu^i$ satisfy the dispersion relation
\begin{equation}
{\rm det} (\lambda I_n-A-\mu B)=0,
\label{disp}
\end{equation}
which defines an algebraic curve of degree $n$ on the $(\lambda, \mu)$-plane.
Moreover,  $\lambda^i$ and $\mu^i$ are required to satisfy the commutativity conditions
\begin{equation}
\frac{\partial_j\lambda^i}{\lambda^j-\lambda^i}=\frac{\partial_j\mu^i}{\mu^j-\mu^i},
\label{comm}
\end{equation}
$i\ne j$, see \cite{Tsarev}. In was observed in \cite{Fer1} that the requirement of the existence of `sufficiently many' hydrodynamic reductions imposes strong restrictions on the system (\ref{2+1}), and provides an efficient classification criterion. To be precise, we will call a system (\ref{2+1}) integrable if, for any $n$,  it possesses infinitely many $n$-component hydrodynamic reductions parametrised by $n$ arbitrary functions of a single variable.
Thus,  integrable systems are required to possess an infinity of $n$-phase  solutions which can be viewed as natural dispersionless analogues of algebro-geometric solutions of soliton equations.

In this Appendix we  illustrate the above procedure by  characterising  all  Hamiltonians $h(v, w)$ for  which the system  (\ref{Hb2}) is integrable by the method of hydrodynamic reductions. Rewriting Eqs. (\ref{Hb2}) as
\begin{equation}
\label{hameq}
\begin{aligned}
v_t &= h_{vw} v_x + h_{ww} w_x, \\
w_t &= h_{vv}v_x + h_{vw}w_x + h_{vw}v_y + h_{ww}w_y,
\end{aligned}
\end{equation}
 we seek n-phase solutions in the form $v = v(R^1, R^2, \dots, R^n), \ w = w(R^1, R^2, \dots, R^n)$ where the phases (Riemann invariants) $R^i$ satisfy the equations
$$
R^i_t = \lambda^i (R) R_x^i, ~~~ R^i_y = \mu^i (R) R_x^i.
$$
The substitution of this ansatz into (\ref{hameq}) implies the relations
\begin{equation}
\label{riem}
\begin{aligned}
v_i\lambda^i &= h_{vw}v_i + h_{ww}w_i, \\
w_i\lambda^i &= h_{vv}v_i + h_{vw}w_i + h_{vw}v_i\mu^i + h_{ww}w_i\mu^i,
\end{aligned}
\end{equation}
here $v_i=\partial_iv, \ w_i=\partial_iw, \ \partial_i=\partial/\partial R^i$. The condition of their non-trivial solvability  implies the dispersion relation for $\lambda^i$ and $\mu^i$,
$$
-{\lambda^i}^2+h_{ww}\lambda^i\mu^i+2h_{vw}\lambda^i+h_{vv}h_{ww}-h_{vw}^2=0.
$$
In what follows we assume that the dispersion relation defines an irreducible conic, which is equivalent to the requirement  $h_{ww}\ne 0, \  h_{vv}h_{ww}-h_{vw}^2 \neq 0$. Setting $v_i = \phi^iw_i$
we can rewrite (\ref{riem}) in the from
\begin{equation}
\label{EqLambdaMu}
\begin{aligned}
 \phi^i\lambda^i &= h_{vw}\phi^i + h_{ww},\\
\lambda^i &= h_{vv}\phi^i + h_{vw}\mu^i\phi^i + h_{ww}\mu^i + h_{vw}.
\end{aligned}
\end{equation}
We also  require the compatibility of  the relations $v_i = \phi^iw_i$, which gives
\be
{\label{SecDW}}\partial_i \partial_j w  = \frac{\partial_j\phi^i}{\phi^j - \phi^i} \partial_iw +  \frac{\partial_i\phi^j}{\phi^i - \phi^j} \partial_jw .
\ee
Expressing $\lambda^i, \mu^i$ in terms of $\phi^i$ from (\ref{EqLambdaMu}),
$$
\lambda^i=\frac{h_{vw}\phi^i+h_{ww}}{\phi^i}, ~~~ \mu^i=\frac{h_{ww}-h_{vv}{\phi^i}^2}{\phi^i(h_{vw}\phi^i+h_{ww})},
$$
and substituting these expressions into the commutativity conditions (\ref{comm}) we obtain relations of the form $\partial_j\phi^i = (\dots)\partial_jw$, $\partial_i\partial_jw = (\dots)\partial_iw\partial_jw$ where dots denote certain rational expressions in $\phi^i, \phi^j$ whose coefficients depend on the Hamiltonian density $h(v,w)$ and its derivatives up to the order three.  The compatibility conditions $\partial_j\partial_k\phi^i = \partial_k\partial_j\phi^i$ take the form $P\partial_jw\partial_kw = 0$ where P is a polynomial in $\phi_i, \phi_j, \phi_k$. Setting all coefficients of this polynomial equal to  zero one obtains the integrability conditions (\ref{4int1}).

\section*{Acknowledgements}

We thank  K Khusnutdinova, O Mokhov and M Pavlov  for clarifying discussions. The research of EVF  was partially supported by the European Research Council Advanced Grant  FroM-PDE.

\end{document}